\documentclass{aa}  
\usepackage{listings}
\usepackage{tabularx} 

\usepackage{graphicx}
\usepackage{caption}
\usepackage{subcaption}
\usepackage{float}
\usepackage{xcolor}
\usepackage{comment}
\usepackage{titlesec}
\usepackage{comment}
\usepackage{mathtools, nccmath}
\usepackage{txfonts}
\begin{document}

   \title{Numerical simulations of MHD jets from Keplerian accretion disks}
   \subtitle{I- Recollimation shocks}  

   \author{T. Jannaud \inst{1}
  \and C. Zanni \inst{2}
  \and J. Ferreira \inst{1}  }

   \institute{Univ. Grenoble Alpes, CNRS, IPAG, 38000 Grenoble, France
\and   INAF - Osservatorio Astrofisico di Torino, Strada Osservatorio 20, Pino Torinese 10025, Italy}

   \date{Received June 21, 2022; accepted October 23, 2022}

 
  \abstract
{The most successful scenario for the origin of astrophysical jets requires a large-scale magnetic field anchored in a rotating object (black hole or star) and/or its surrounding accretion disk. Platform jet simulations, where the mass load onto the magnetic field is not computed by solving the disk vertical equilibrium but is imposed as a boundary condition, are very useful to probe the jet acceleration and collimation mechanisms. The drawback is the very large parameter space: despite many previous studies, it is very difficult to determine the generic results that can be derived from them.}
%
{We wish to establish a firm link between jet simulations and analytical studies of magnetically-driven steady-state jets from Keplerian accretion disks. In particular, the latter have predicted the existence of recollimation shocks due to the dominant hoop-stress, so far never observed in platform simulations.}  
%
{We perform a set of axisymmetric MHD simulations of non-relativistic jets using the PLUTO code. The simulations are designed to reproduce the boundary conditions generally expected in analytical studies. We vary two parameters: the magnetic flux radial exponent $\alpha$ and the jet mass load $\kappa$. In order to reach the huge unprecedented spatial scales implied by the analytical solutions, a new method allowing to boost the temporal evolution has been used.}
%
{We confirm the existence of standing recollimation shocks at large distances, behaving qualitatively with the mass load $\kappa$ as in self-similar studies. The shocks are weak and correspond to oblique shocks in a moderately high fast-magnetosonic flow. The jet emitted from the disk is focused towards the axial inner spine, which is the outflow connected to the central objet. The presence of this spine is shown to have a strong influence on jet asymptotics. We also argue that steady-state solutions with $\alpha \geq 1$ are numerically out of range.} 
%
{Internal recollimation shocks may produce observable features such as standing knots of enhanced emission and a decrease of the flow rotation rate. However, more realistic simulations, e.g. fully three-dimensional, must be done in order to investigate non-axisymmetric instabilities and with ejection only from a finite zone in the disk, so as to to verify whether these MHD recollimation shocks and their properties are maintained.}


   \maketitle
%

\section{Introduction}

Astrophysical jets are commonly observed in most, if not all, types of accreting sources. They are emitted from young stellar objects \citep{bally2007,ray2007,ray2021}, active galactic nuclei and quasars \citep{boccardi2017}, close interacting binary systems \citep{fender2014, tudor2017} and even post-AGB stars \citep{bollen2017}. Despite the different central objects (be it a black hole, a protostar, a white dwarf or a neutron star), these jets share several properties: (i) they are supersonic collimated outflows with small opening angles, (ii) the asymptotic speeds scale with the escape speed from the potential well of the central object and (iii) they carry away a sizable fraction of the power released in the accretion disk. Since the only common feature shared by all these different astrophysical objects is the existence of an accretion disk, it is natural to seek for a jet model that is related to the disk and not to the central engine. This universal approach is further consistent with the accretion-ejection correlations observed in these objects (see e.g. \citealt{merloni2003,corbel2003,gallo2004,coriat2011,ferreira2006, cabrit2007} and references therein).         

%

Despite these general common trends, astrophysical jets do show some differences in their collimation properties. For instance, the core-brightened extragalactic jets, classified as FRI jets \citep{fanaroff1974}, appear conical and show large-scale wiggles (see e.g. \citealt{laing2013, laing2014} and references therein). On the opposite, the edge-brightened FRII jets appear nearly cylindrical and with a terminal hotspot \citep{laing1994,boccardi2017}. Most of the jets imaged with very long baseline interferometry do not appear as continuous flows, but can be modeled as a sum of discrete features, known as blobs or knots, usually associated with shocks \citep{zensus1997}. Those shocks are assumed to originate either from pressure mismatches at the jet boundary with the external medium or from major changes at the base of the flow (e.g., new plasma ejections or directional changes), with some of these knots being stationary features (e.g. \citealt{lister2009,lister2013, walker2018, doi2018, park2019}). 

On the other hand, jets from young forming stars do not seem to have such a clear FRI/FRII dichotomy and often display evidences of a conflictual interaction (shocks) with the ambient cloud medium \citep{reipurth2001}. This might be consistent with the suspicion that the FR dichotomy would only be a consequence of the jet interaction with its environment, with low-power jets remaining undisrupted and forming hotspots in lower mass hosts \citep{mingo2019}. However, protostellar jets might also have intrinsic collimation properties different from those of extragalactic jets, possibly because they are non relativistic outflows. This is an open question, unsettled yet.


Since the seminal model of \citet{blandford1982} (hereafter BP82), it is known that a large scale vertical magnetic field threading an accretion disk is capable of accelerating the loaded disk material up to super fast-magnetosonic speeds. This acceleration, usually termed magneto-centrifugal, goes along with an asymptotic collimation of the ejected plasma, thanks to the magnetic tension associated with the toroidal magnetic field (hoop-stress). In this semi-analytical model, a self-similar ansatz has been used allowing to solve the full set of stationary ideal magnetohydrodynamic (hereafter MHD) equations. Later, this self-similar jet model has been generalized in different ways by playing with the magnetic field distribution \citep{contopoulos1994,ostriker1997}, thermal effects \citep{vlahakis2000,ceccobello2018} and even extended to the relativistic regime 
\citep{li1992,vlahakis2003,polko2010, polko2014}. However, it is unclear if self-similarity affects the overall jet collimation properties. Not only are the axis and the jet-ambient medium region both not taken into account, but the final outcome of the jets (i.e. acceleration efficiency, jet kinematics and opening angle, presence of radial oscillations or even shocks) may well be also impacted by the imposed geometry. 
 
Using the only class of self-similar jet models smoothy connected to a quasi-Keplerian accretion disk, \citet{ferreira1997} (hereafter F97) showed that these super fast-magnetosonic jets systematically undergo a refocusing toward the axis (see also \citealt{polko2010}). Such a recollimation is due to the dominant effect of the internal hoop-stress and has nothing to do with a pressure mismatch at the jet-ambient medium interface, proposed to explain knotty features in extragalactic jets \citep{komissarov1998,perucho2007,perucho2020}. According to F97, recollimation would be generic to MHD jets anchored over a large range of Keplerian accretion disks. This is indeed verified for warm outflows \citep{casse2000} and weak magnetic fields \citep{jacquemin-ide2019}.

While MHD recollimation is also seen in non self-similar works (e.g. \citealt{pelletier1992}), other strong assumptions are usually made leaving the question of the jet asymptotics open. \citet{heyvaerts1989} used another approach based on the electric poloidal current (or Poynting flux) still present at infinity. They showed that any stationary axisymmetric magnetized jet will collimate at large distances from the source to paraboloids or cylinders, depending if the asymptotic electric current vanishes or not. This important theorem has been later generalized \citep{heyvaerts2003} by taking into account the issue of current closure and its effect on the geometry of the solution \citep{okamoto2001,okamoto2003}. However, the theorem only addresses the {\em asymptotic} electric current and since no simple connection with the source can be made, it is unclear how much current is actually left asymptotically.

Connecting the asymptotic electric current to the source is naturally done with time-dependent MHD simulations. Those reaching the largest spatial scales treat the accretion disk as a boundary condition, allowing the jet dynamics to be studied independently of the disk \citep{ustyugova1995, ouyed1997,ouyed1997a,ouyed1999, ustyugova1999, krasnopolsky1999, krasnopolsky2003, ouyed2003, anderson2005, anderson2006, fendt2006,pudritz2006, porth2010, porth2015, staff2010, staff2015, stute2014, barniolduran2017, tesileanu2014, tchekhovskoy2016, ramsey2019}. The drawback of these platform jet simulations is their huge degree of freedom, since several distributions must be specified at the lower injection boundary. It therefore has been very difficult to determine the exact generic results on jet collimation that can be derived from them.  

To summarize, despite many theoretical and numerical studies, no connection has been firmly established between the jet launching conditions and the jet collimation properties at observable scales. This work is the first of a series aiming at bridging this gap. Our approach here is to assess whether the general results obtained within the self-similar framework still holds in full 2D time-dependent simulations. We will address in particular whether the existence of recollimation shocks is indeed unavoidable for the physical conditions expected in Keplerian accretion disks, as proposed by F97. 

As a consequence of this approach, we will focus only on steady-state jets, allowing us to directly confront our simulations with MHD jet theory. It is clear that most if not all astrophysical jets exhibit time-dependant features, see e.g. \cite{cheung2007} for M87 or \cite{bally2007} for young stars. However, our goal is not to reproduce a specific astrophysical jet but instead to deduce generic behaviours of MHD jets emitted from keplerian accretion disks.

The paper is organized as follows. Section~2 describes our numerical setup and boundary conditions mimicking an axial spine (related to the central object) surrounded by a self-similar cold jet. Since analytical studies require huge spatial and temporal scales, a special temporal numerical scheme has been designed. Our reference simulation, which corresponds to a typical BP82 jet, is described in length in Section~3. It is shown that recollimation shocks are indeed obtained in agreement with the analytical theory. It is the first time that such shocks are obtained self-consistently, showing that these are not artificial biases due to the mathematical ansatz used but consequences of the jet launching conditions. A parametric study is made in Section~4, where we vary the magnetic flux exponent $\alpha$ and the jet mass load $\kappa$, confirming the striking qualitative correspondance between our numerical simulations and analytical solutions. In particular, we show that the asymptotic jet collimation depends mostly on the exponent $\alpha$. However, the existence of an axial spine introduces quantitative differences hinting towards to a possible role of the central object in affecting the collimation properties of the jets emitted by the surrounding disk. Our results are finally confronted to the wealth of previous 2D numerical simulations in Section 5 and we conclude in Section 6.

\section{MHD simulations of jets from Keplerian disks}

\subsection{Physical framework and governing equations}

\begin{figure}
\centering
  \includegraphics[width=.8\linewidth,trim=5 5 5 5, clip]{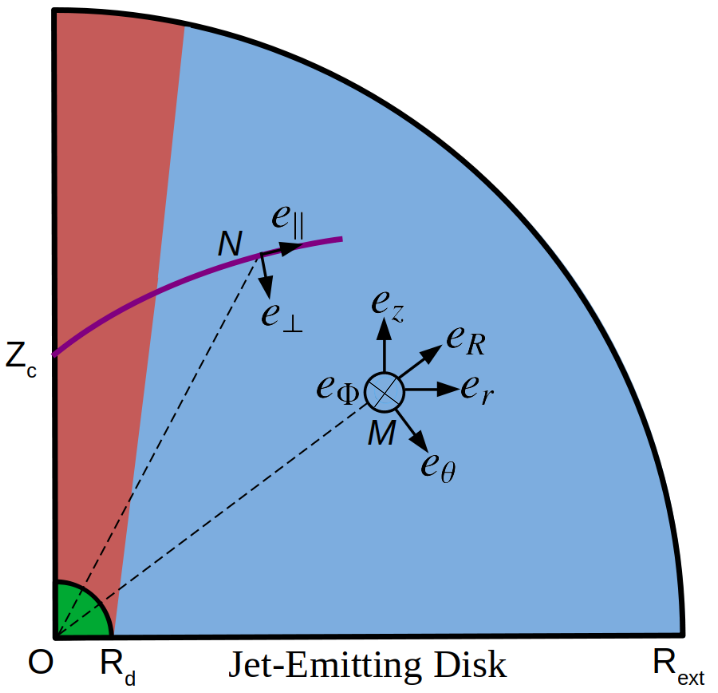}
\caption{Sketch of our computational domain. The central object and its interaction with the innermost disk are located below the inner boundary at $R_d$ (green region), the near Keplerian Jet-Emitting Disk (JED) being established from $R_d$ to the end of the domain $R_{ext}$. An axial outflow (the spine) is emitted from the central regions (in red) and the jet is emitted from the JED (in blue). The solid purple line represents a recollimation shock surface starting on the axis at an height $Z_{c}$. For each point N lying on this surface, we will use local poloidal unit vectors $(\vec{e_\perp}, \vec{e_\parallel})$, respectively perpendicular and parallel to the shock surface. Also, at any point $M$ inside the domain, we will either use spherical ($\vec{e_R}$,$\vec{e_\theta}$,$\vec{e_\phi}$) or cylindrical ($\vec{e_r}$,$\vec{e_\phi}$,$\vec{e_z}$) coordinates.
}
\label{fig:SchemaSetup}
\end{figure}

We intend to study the collimation properties of magnetically driven jets emitted from Keplerian accretion disks, as depicted in Figure~\ref{fig:SchemaSetup}. The disk is settled from an inner radius $R_d$ to an outer radius $R_{ext}= 5650.4 R_d$ and is assumed to be orbiting around a central object of mass $M$ located at the center of our coordinate system. The disk itself is not computed and we assume that it behaves like a Jet-Emitting Disk (hereafter JED), with consistent prescribed boundary conditions. Since we use a spherical grid, the central object as well as its interaction with the disk are assumed to occur inside a sphere of radius $R_d$ (the green zone in Fig.~\ref{fig:SchemaSetup}). That sphere defines the inner boundary that will be discussed below. 

We further assume that a large scale magnetic field is threading both the disk and the central object. The existence of this field allows the production of two outflows, one from the disk (blue region in Fig.~\ref{fig:SchemaSetup}) and one from the central spherical region (red region in Fig.~\ref{fig:SchemaSetup}). Hereafter, we will always refer to the disk emitted outflow as the "jet", and to the outflow emitted from the spherical region (green zone in Fig.~\ref{fig:SchemaSetup}) as the "spine". Since our goal is to focus on the dynamics of the jet itself, we will try to limit as much as possible the influence of the spine. 

Two systems of coordinates centered on the mass $M$ will be used, spherical $(R,\theta,\phi)$ and cylindrical $(r,\phi,z)$. Both the spine and the jet are assumed to be in ideal MHD and we numerically solve the usual set of MHD equations. This includes mass conservation  
\begin{equation}
\frac{\partial \rho}{\partial t} + \nabla \cdot (\rho \vec{u}) =0 ,
\end{equation}
where $\rho$ is the density and ${\bf u} $ the flow velocity, 

the momentum equation 
\begin{equation}
\frac{\partial \rho {\bf u}}{\partial t} + \nabla \cdot \left [ \rho {\bf u}{\bf u} + \left ( P + \frac{{\bf B}\cdot{\bf B} }{2\mu_o}\right) - \frac{{\bf B}{\bf B}}{\mu_o} \right ] =- \rho \nabla \Phi_G 
\end{equation}
where $P$ is the thermal pressure, ${\bf B}$ the magnetic field and $\Phi_G=-GM/R$ the gravitational potential due to the central mass.

The evolution of the magnetic field is determined by the induction equation
\begin{equation} 
\frac{\partial \vec{B}}{\partial t} + \nabla  \times (\vec{B} \times \vec{u}) = 0
\end{equation}
Since we will focus on highly supersonic flows, we decided to derive the pressure $P$ and internal energy by solving the entropy equation 
\begin{equation}
\frac{\partial \rho S}{\partial t} + \nabla \cdot \left( \rho S \vec{u} \right) = 0
\end{equation} 
where $S = P/\rho^\Gamma$ is the specific entropy and $\Gamma= 1.25$ is the polytropic index (the same for all our simulations). This simple advection equation guarantees that the pressure will not assume nonphysical (e.g. negative) values. But on the other hand, it will fail at providing the correct entropy jump in correspondence of shocks. However, as long as the thermal energy of the flow remains negligible compared to the kinetic and magnetic energy, this should not represent an issue. 

Thanks to axisymmetry, the poloidal magnetic field can be computed using the magnetic flux function $\Psi$ (which corresponds to $R\sin\theta \, A_\phi$, where $\vec{A}$ is the vector potential), 
\begin{equation}
B_R = \frac{1}{R^2 \sin\theta} \frac{\partial \Psi}{\partial \theta} \qquad
B_\theta = -\frac{1}{R\sin\theta}\frac{\partial \Psi}{\partial R}
\end{equation}
which already verifies $\nabla\cdot \vec{B}=0$. An axisymmetric magnetized jet can therefore be seen as a bunch of poloidal magnetic surfaces defined by $\Psi (R,\theta)= constant$, nested around each other and anchored on the disk for the jet and in the central object for the spine.  

In steady-state, equations (1) to (4) lead to the existence of the following five MHD invariants, namely quantities that remain constant along each magnetic surface (\cite{weber1967}):    
\begin{itemize}
\item the mass flux to magnetic flux ratio $\eta (\Psi) = \mu_o \rho u_p/B_p$,
\item the rotation rate of the magnetic surface  $\Omega_* (\Psi) = \Omega - \eta B_\phi/(\mu_0 \rho r)$,
\item the total specific angular momentum carried away by that surface $L(\Psi) = \Omega r^2 - rB_\phi/\eta$,
\item the Bernoulli invariant $E(\Psi) = \frac{u^2}{2} + H +\Phi_G - \Omega_{*}r B_{\phi}/\eta $,
\item the specific entropy $S(\Psi)=P/\rho^\Gamma$,
\end{itemize}
where $\Omega= u_\phi/r$ and $H=\frac{\Gamma}{\Gamma-1}\frac{P}{\rho}$ is the specific enthalpy. We will make use of these relations when designing boundary conditions.

\subsection{Numerical setup}

We solve the above set of equations using the MHD code PLUTO\footnote{PLUTO is freely available at http://plutocode.ph.unito.it} \citep{mignone2007}. 
We configured PLUTO to use a second-order linear spatial reconstruction with a monotonized-centered limiter on all the variables. This method provides the steeper linear reconstruction compatible with the stability requirements of the scheme. A flatter and more diffusive linear reconstruction is employed in a few cells around the rotation axis to damp numerical spurious effects that typically appear in these zones due to the discretization of the equation around the axis's geometrical singularity. The HLLD Riemann solver of \citet{miyoshi2005} is employed to compute the intercell fluxes. This solver is one of the best suited to properly capture Alfv\'{e}n waves, a crucial element to properly model trans-Alfv\'{e}nic flows. So as to match the order of the spatial reconstruction, we chose a second-order Runge-Kutta scheme to advance the equations in time. The $\nabla\cdot \vec{B}=0$ condition is ensured employing a Constrained Transport (CT) scheme, enforcing that constraint at machine accuracy. 

The two-dimensional computational domain is discretized using spherical coordinates $(R,\theta)$ assuming axisymmetry around the rotation axis of the disk. The domain encompasses a spherical sector going from the polar axis ($ \theta= 0$) to the surface of the disk, assumed to be $\theta = \pi/2$ for simplicity, resolved with $N_\theta= 266$ points in the $\theta$ direction. The cell size in the $\theta$ direction is mostly uniform, but decreases on a few cells near the axis. This essential in our setup, as the expected collimation shocks are formed near the axis: a too low resolution in this zone would prevent their formation.
In the radial direction,  the grid goes from $R_d$ to $R_{ext}= 5650.4 R_d$ with $N_R=1408$ points in a logarithmic spacing ($\Delta R \propto R$) so as to keep cells approximately squared cells $(\Delta R \approx R\Delta \theta)$ far from the axis.

We choose such a huge numerical domain because our goal is to capture the recollimation shocks predicted in the self-similar solutions of F97. According to his Fig.~6, those shocks may occur at altitudes spanning from several hundreds to a thousand times the jet launching radius. Our spherical grid with a ratio of $5650.4$ therefore provides a suitable range to observe such shocks. The drawback of these huge spatial scales is of course the terrible contrast in time scales. The Keplerian time scaling in $r^{3/2}$ imposes that in order to compute a full orbit at the outer disk edge, the inner one should have done over $4.10^5$ orbits. This would be hardly affordable if we were to use a standard evolutionary scheme. In order to achieve such long time scales, we designed a specific method that accelerates the numerical integration by using larger and larger time steps to evolve the equations as the solution starts to converge towards a steady-state.
This method is very successful and allowed us to significantly boost the evolution of our jets (see Appendix A).        

\subsection{Initial conditions}

Our initial magnetic field is assumed to be potential, which  leads to a second order partial differential equation on $\Psi(R,\theta)$. In order to represent suitable self-similar solutions, we solve this equation by assuming 
\begin{equation}
\Psi= \Psi_d (R/R_d)^\alpha \Phi(\theta) \,  
\end{equation} where the function $\Phi\left(\theta\right)$ has been determined assuming that the initial field is potential, i.e. current-free and force-free ($J_\phi = 0$).

The exponent $\alpha$ is a free parameter of the model leading to $B_R \propto B_\theta \propto R^{\alpha-2}$. For $\alpha=0$, field lines are conical, for $\alpha=1$ they are parabolic and $\alpha=2$ describes a constant (straight) vertical field. The seminal BP82 solution is for $\alpha=3/4$. 
    
The magnetic field being potential, no magnetic force is imposed initially on the plasma. It is therefore assumed to be in a spherical symmetric hydrostatic equilibrium (${\bf u}=0$) with $dP/dR= - \rho GM/R^2$. We choose the following trivial solution 
\begin{equation}
\begin{aligned}
\rho & = \rho_{a} \left(\frac{R}{R_d}\right)^{2\alpha-3} \\
P    & = \frac{1}{4-2\alpha}\frac{\rho_a GM}{R_d}\left(\frac{R}{R_d}\right)^{2\alpha-4} \, ,
\end{aligned}
\end{equation}
The sound speed $C_s$ is defined as $C_s^2 = \partial P/\partial \rho = \Gamma P/\rho$. In the following, $\rho_a$ will be referred to as the density at the axis, right above the central sphere.

\subsection{Boundary conditions}

Boundary conditions must be imposed at the polar axis ($\theta=0$), the outer frontier ($R=R_{ext}$), at the JED surface ($\theta=\pi/2$, $R$ from $R_d$ to $R_{ext}$) and at the spine boundary ($R=R_d$, $\theta$ from 0 to $\pi/2$).  
On the polar axis, usual proper reflecting boundary conditions are imposed on all quantities.
The special treatment done for the other three boundaries is described, especially for the JED and spine boundaries where mass is being injected.

\subsubsection{Outer boundary ($R=R_{ext}$)}

At the outer frontier "outflow" conditions are imposed : For $\rho$, $P$, $B_R$, $B_\theta$, $R B_\phi$, $u_R$, $u_\theta$ and $u_\phi$, the gradient along the radial direction is conserved, and we use the Van Leer slope limiter to avoid spurious oscillations. Additionally, we enforce a positive toroidal Lorentz force on the subalfv\'enic part of this boundary.

\subsubsection{Jet generation: the Jet Emitting Disk ($\theta=\pi/2$)}

We need to specify eight quantities ($\rho, {\bf u}, {\bf B}, P$) that must be representative of the fields expected at the surface of a JED. As the lifted material gets accelerated along a field line, its poloidal velocity will become larger than the slow magnetosonic $V_{sm}$, poloidal Alfv\'en $V_{Ap}$ and fast magnetosonic $V_{fm}$ phase speeds. Crossing each of these critical speeds defines a regularity condition that determines one quantity at the jet basis, leaving therefore 5 free functions to specify. However, we wish to control the mass loss from the JED, which requires that the injected outflow is already super-slow magnetosonic (hereafter super-SM). We therefore have to impose six functions at the JED boundary, leaving two free to adjust over time, $B_\phi$ and $B_R$, the latter controlling the magnetic field bending.  

Our choice of boundary conditions at $\theta=\pi/2$ (so that $R=r$) is then the following 
\begin{equation}
\begin{aligned}
\rho &= \rho_d \left(\frac{R}{R_d} \right)^{2\alpha-3} \\
P &= \rho_d \frac{C_{s_d}^2}{\Gamma} \left(\frac{R}{R_d} \right)^{2\alpha-4} \\
B_\theta & = -B_d \left(\frac{R}{R_d} \right)^{\alpha-2} \\
u_\theta & = -u_d \left(\frac{R}{R_d} \right)^{-1/2} \\
u_R & = u_\theta \frac{B_R}{B_\theta} \\
u_\phi & = \Omega_* r\, +\,  u_\theta\frac{B_\phi}{B_\theta}
\end{aligned}
\label{eq:InjectionCondition}
\end{equation}
where $\Omega_*$ is the angular velocity of the magnetic surfaces (an MHD invariant in steady-state). We assume $\Omega_*= \Omega_K= \sqrt{GM/r^3}$, in agreement with a near Keplerian accretion disk, leaving four normalizing quantities $\rho_d, C_{s_d}, B_d$ and $u_d$ to  be specified at $R_d$. These distributions are consistent with a self-similar JED and describe an ideal steady MHD flow with ${\bf u}_p \parallel {\bf B}_p$\footnote{With this condition the $\phi$ component of the electric field $\vec{E} = -\vec{u}\times\vec{B}$ is zero and the magnetic flux distribution does not change in time.}, anchored on a disk that imposes magnetic field lines rotating at the Keplerian angular velocity $\Omega_K$. Note that the fixed component of the magnetic field threading the disk ($B_\theta$) is actually the initial condition to conserve the magnetic flux injected in the computational domain, and only $B_R$ is allowed to vary in response to the jet dynamics.  

In order to pick up values at $R_d$ that are consistent with the jet calculations done by BP82 or F97, we express the JED boundary conditions as a function of four dimensionless parameters: 
(1) the jet density $\rho_d$ is fixed with respect to the density at the polar axis using $\rho_d= \delta \rho_a$;
(2) the disk sound speed (temperature) is defined relatively to the Keplerian speed with $\epsilon= C_{sd}/V_{Kd}$;
(3) the magnetic field strength $B_d$ is controlled by measuring the $\theta$ component of the poloidal Aflv\'en speed with respect to the Keplerian speed, namely $\mu = V_{Ad}/V_{Kd}= B_d\sqrt{R_d/(\mu_o\rho_d GM)}$;
(4) the (vertical) injection velocity \textbf{$u_d$} can be determined with the well known mass loading parameter $\kappa$ introduced by BP82 using 
\begin{equation}
\kappa= \frac{\mu_o \rho_d u_d V_{Kd}}{B_d^2} = \frac{u_d V_{Kd}}{V_{Ad}^2} = \frac{u_d}{V_{Kd}} \frac{1}{\mu^2} \ .
\end{equation}
By fixing $u_d/V_{Kd} = 0.1$ for all the simulations we obtain $\kappa=0.1/\mu^2$. In order to be able to fix the value of the injection speed $u_d$ and therefore the JED mass flux, we must require that $u_p > V_{sm}$. Since we are mostly interested in producing cold MHD outflows, we assume $\epsilon = 0.01$ so that the $\theta$ component of the sonic Mach number is $M_{s\theta} = u_d/C_s =  u_d/V_{Kd}\epsilon = 10$. Since the total poloidal speed at the jet boundary is larger than $u_d$, the sonic Mach number $M_s = u_p/C_s > 10$. 
Since the poloidal Alfv\'en speed at the disk surface is much larger than the sound speed, $C_s > V_{sm}$ and $M_s >1$ is enough to warrant a super-SM condition. 

We decided to vary the mass load and the disk Alfv\'en speed by only changing the disk density $\rho_d$ (and keeping the injection speed $u_d$ and the disk magnetic field $B_d$ constant for all the simulations). As a consequence, the density contrast $\delta$ can be expressed as a function of $\mu$ (or $\kappa$). We assumed the relation $\delta = 100 /\mu^2 = 1000 \kappa$. Notice that with our parametrization the JED boundary conditions are determined by one dimensionless parameter only, typically $\kappa$, while the other two free parameters $\mu$ and $\delta$ are determined as a function of $\kappa$, and $\epsilon$ is fixed for all the simulations.

\subsubsection{Spine generation: the central object ($R=R_d$)}
 
In the spine we follow a similar methodology as in the JED and specify six quantities along the inner spherical boundary at $R_d$. This leaves again two quantities that are free to evolve, $B_\phi$ and $B_\theta$. In order to conserve the magnetic flux injected in the computational domain, we fix $B_R(\theta)$ to its initial value. Notice that since the $B_\theta(R)$ profile is fixed along the JED boundary ($\theta = \pi/2$) and $B_R(\theta)$ is kept constant in time along the spine boundary ($R=R_d$), the total poloidal field and its inclination $B_R/B_\theta$ does not change with time at the inner radius of the disk ($R=R_d, \theta = \pi/2$).
The strength of the magnetic field is already determined by the value of $\mu$ chosen in the JED. Since the outflowing material leaving the central region is in ideal MHD and we are looking for a steady jet, one has $u_\theta  = u_R B_\theta/B_R$. This leaves us with the four distributions $\rho,  C_s, u_R$ and $u_\phi$ to be specified along $\theta$.
    
If the central object possesses its own magnetosphere, then $R_d$ might be considered as the disk truncation radius. What would be encapsulated within $R_d$ could then be a complex combination of a stellar wind plus any type of magnetospheric wind (steady or not, see for instance \citealt{zanni2013} and references therein). If the central object is instead a black hole, then $R_d$ might be considered as the innermost stable circular orbit and what is hidden inside $R_d$ would highly depend on the black hole spin. While a non rotating black hole would provide no outflow, a rather strong magnetic flux concentration is seen to occur in GRMHD simulations of spinning black holes, leading to the generation of powerful outflows through the Blandford-Znajek processs (see e.g. \cite{blandford1977,tchekhovskoy2010,liska2018} and references therein). 

However our goal is to study the outcome of the jet emitted from the disk. We thus decided to minimise as much as possible the influence of the spine. It turned out to be an almost impossible task but this will be seen later on. As pointed out in early works on magnetized rotating objects (e.g. \citealt{ferreira1995}), the jet power depends on the available electromotive force (hereafter emf) $e= \int {\bf E}_m\cdot {\bf dl} = \int ({\bf u}\times {\bf B}_p)\cdot {\bf dl}$. While the disk provides an emf $e_{disk} \simeq \int \Omega_K r B_z dr$, the central region provides $e_{obj}\simeq \int \Omega r B_R R_d d\theta$. An obvious way to decrease $e_{obj}$ is thus to allow $\Omega$ to decrease as one goes from the disk to the pole. 
We thus use (in agreement with steady-state ideal MHD) $u_\phi = \Omega_* r\, +\,  u_R B_\phi/B_R$, with magnetic surfaces rotating as 
\begin{equation}
\Omega_* = \Omega_a (1-f(\theta)) + \Omega_{Kd}f(\theta)
\end{equation}
where $f(\theta)$ is a spline function varying smoothly from zero at $\theta=0$ to unity $\theta =\pi/2$ (see Appendix B). Most of the simulations presented in this paper were done with $\Omega_a =0$ (but see Sect.~\ref{sec:spine}). This choice is consistent with a non-rotating black hole but also with an innermost disk radius (our $R_d$) well below the co-rotation radius in the case of a star. 

The fixed radial speed is defined through the sonic Mach number $M_{sR}$, $u_R = M_{sR} C_s$. For $M_{sR}$ we assume a constant value along $\theta$ that can be derived from the JED boundary conditions by assuming its continuity at the inner disk radius $R_d$, $M_{sR} = M_{s\theta} \lvert B_R/B_\theta \rvert_d = 10 \lvert B_R/B_\theta \rvert_d > 1$. Since the field inclination at the inner disk radius $\lvert B_R/B_\theta\rvert_d $ is constant, also $M_{sR}$ does not change with time.
The sound speed at the base of the spine is computed as 
\begin{equation}
C_s = C_{sa} (1-f(\theta)) + C_{sd}f(\theta)
\end{equation}
where the sound speed on the axis $C_{sa}$ is computed so as to verify the Bernoulli integral $E_a= E(\theta=0)$ at the axis. Since the MHD contribution vanishes on the axis, one gets directly  
\begin{equation}
C_{sa}^2 = \frac{GM}{R_d} \frac{1 + e_a}{\frac{1}{2}M_{sR}^2+\frac{1}{\Gamma-1}}
\end{equation}
and
\begin{equation}
\label{eq:ur}
u_{Ra}^2 = \frac{GM}{R_d} \frac{1 + e_a}{\frac{1}{2}+\frac{1}{M^2_{sR}\left(\Gamma-1\right)}} \ ,
\end{equation}
where $u_{Ra}$ is the injection radial speed on the axis and $e_a=  E_aR_d/GM$ is the Bernoulli integral normalized to the gravitational energy at $R_d$ and will be used as a parameter to fix the axial spine temperature. Note that the normalized Bernoulli integral for the jet at $R_d$ writes $e_d = \lambda_d - 3/2 + \epsilon^2/(\Gamma-1)$, where $\lambda = L(\Psi)/\sqrt{GMr_o}$ is the magnetic lever arm parameter, measured here at the anchoring radius $r_o=R_d$. Since our jets are cold, enthalpy plays no role and $e_d$ is mostly determined by $\lambda$ (which is known only once the simulation has converged to a steady-state). For our simulations, we expect $\lambda$ around 10 (see our parameter space Fig.~\ref{fig:LambdaKappa}). We thus fix $e_a=2$ in order to obtain a spine with a smaller energetic content than the surrounding jet. Notice that with our choice of parameters the injection speed along the axis, Eq.(\ref{eq:ur}), is higher than the escape speed. Since the flow is cold and there is no magneto-centrifugal acceleration along the symmetry axis, the flow will gradually slow down along $R$ in the spine from this very high speed in its core. The spine flow can cross the Alfv\'en and fast-magnetosonic critical points due to a decrease of the magnetic field intensity, not thanks to a flow acceleration.

Finally for the density, we need to smoothly connect its axial value $\rho_a$ to the much larger value injected at the disk surface $\rho_d$. We choose to do this by computing $\rho (\theta)= \eta B_R/(\mu_o u_R)$, with the MHD invariant $\eta$ following  
\begin{equation}
\eta = \eta_{a} (1-f(\theta)) + \eta_{d}f(\theta)
\end{equation}
$\eta_{a}$ and $\eta_{d}$ being fully determined (see Appendix B). This method ensures that the mass flux to magnetic flux ratio has a smooth variation from the disk to the axis. For numerical stability reasons, as the strongest magnetic field is on the axis, the density in the code is normalized to $\rho_a$, providing a dimensionless density at the axis of 1.

\subsection{Summary of parameters and normalisation} 
\label{sec:summary}

\renewcommand{\arraystretch}{1.2}
\begin{table*}[h]
\centering
\begin{tabular}{ |p{0.8cm}|p{0.65cm}|p{0.8cm}|p{0.7cm}|p{0.6cm}|p{0.65cm}|p{0.8cm}|p{1.15cm}|p{1.1cm}|p{0.7cm}|p{0.7cm}|p{0.8cm}|p{0.65cm}|p{0.7cm}|  }
\hline
    Name & $\kappa$ & $\alpha$ & $\mu$ & $\delta$ & $\frac{t_{end}}{10^5}$ & $Z_{shock}$ & $\theta^{ext}_{FM}(rad)$ & $\theta^{ext}_{A}(rad)$ & $r_{o,FM}$ & $\dot{M}_{jet}$ &  $\frac{\dot{M}_{spine}}{\dot{M}_{jet}}$ & $P_{jet}$ & $\frac{{P}_{spine}}{{P}_{jet}}$\\
\hline
\hline
    K1 & 0.05 & 12/16 & 1.41 & 50 & 7.34 & 2150 & 0.64 & 0.94 & 301 & 179 & 0.102 & 492 & 0.82\\
\hline
    K2 & 0.1 & 12/16 & 1.00 & 100 & 6.51 & 1850 & 0.67 & 1.05 & 323 & 363 & 0.096 & 616 & 0.81\\
\hline
    K2l & 0.1 & 12/16 & 1.00 & 100 & 12.3 & 2490 & 0.65 & 1.02 & 289 & 357 & 0.094 & 620 & 0.81\\
\hline
    K3 & 0.2 & 12/16 & 0.71 & 200 & 10.1 & 1810 & 0.69 & 1.09 & 368 & 743 & 0.093 & 768 & 0.80\\
\hline
    K4 & 0.5 & 12/16 & 0.45 & 500 & 8.67 & 1150 & 0.90 & 1.26 & 655 & 2040 & 0.093 & 1024 & 0.81\\
\hline
    K5 & 1.0 & 12/16 & 0.32 & 1000 & 4.62 & 700 & 0.99 & 1.34 & 670 & 4095 & 0.116 & 1264 & 0.96\\
\hline
    A1 & 0.1 & 10/16 & 1.00 & 100 & 9.08 & 1900 & 0.96 & 1.23 & 234 & 195 & 0.206 & 551 & 1.21\\
\hline
    A2 & 0.1 & 11/16 & 1.00 & 100 & 8.34 & 1800 & 0.87 & 1.15 & 349 & 272 & 0.137 & 578 & 0.99\\
\hline
    A3 & 0.1 & 13/16 & 1.00 & 100 & 5.79 & 1920 & 0.59 & 0.95 & 566 & 690 & 0.047 & 668 & 0.66\\
\hline
    A4 & 0.1 & 14/16 & 1.00 & 100 & 6.26 & 2050 & 0.64 & 0.94 & 398 & 1321 & 0.023 & 740 & 0.53\\
\hline
    A5 & 0.1 & 15/16 & 1.00 & 100 & 1.62 & 2030 & 0.50 & 0.83 & 1046 & 3275 & 0.009 & 848 & 0.41\\
\hline
    SP  & 0.1 & 12/16  & 1.00 & 100 & 3.93 & 1250 & 0.82 & 1.09 & 506 & 392 & 0.097 & 613 & 0.98\\
\hline
\end{tabular}
\caption{List of simulations presented in this paper. All the simulations presented have been performed in the grid described in Section 2.2 (e.g. $N_R=1408$ and $N_\theta=266$) except K2l, performed in a lower resolution grid (e.g. $N_R=704$ and $N_\theta=144$) The parameters $\kappa$ and $\alpha$ are varied independently allowing to compute $\mu$, $\delta$ and $M_S$. The columns $Z_{shock}, \theta^{ext}_{FM}, \theta^{ext}_{A}, r_{o,FM}, \dot M_{jet}, \dot M_{spine}/\dot M_{jet}, P_{jet}$ and $P_{spine}/ P_{jet}$ are outputs of the simulation measured at the final time $t_{end}$ (given in $10^5 T_d$ units). Simulations K1 to A5 were done with a non rotating spine, namely $\Omega_a=0$ and $e_a=2$. Simulation SP is done for $\Omega_a=\Omega_{Kd}$ and $e_a=10$ and is discussed in Sect~\ref{sec:spine}. 
See Sect.~\ref{sec:summary} for the definition of all these quantities.}
\label{tab:ParametresSimus}
\end{table*}

Each simulation is entirely determined by the following dimensionless parameters and quantities:  
\begin{itemize}
\item the radial exponent $\alpha$ of the magnetic field
\item the Bernoulli parameter $e_a$ (equal to 2 for most cases) and the spine angular frequency on the axis $\Omega_a$ (equal to 0 in most simulations).   
\item the cold jet parameters: $\kappa$, $\mu = \sqrt{0.1/\kappa}$, $\delta = 1000\kappa$, $\epsilon=0.01$
\end{itemize}
With our choices, we ensure that the injected flow is everywhere super-SM and that the main emf is due the JED which is magnetically launching a cold jet. The profiles of several quantities along the magnetic flux near the lower boundary (inner spherical boundary and disk) at the final time reached by our reference simulation K2 can be seen in Fig.~(\ref{fig:Limit conditions}).  

This leaves us with only two free parameters, $\alpha$ and $\kappa$. We will not explore here their whole range but keep them within the parameter space of jets from JEDs as obtained by F97 but also by the solutions of \citet{contopoulos1994} and the simulations of \citet{ouyed1997}. The radial exponent $\alpha$ is varied from $10/16$ to $15/16$. In a strict self-similarity, this exponent must be consistent with the underlying disk, namely $\alpha= (12+8\xi)/16$, where $\xi$ is the disk ejection efficiency defined with the disk accretion rate as $\dot M_a(r) \propto r^\xi$ \citep{ferreira1995}. However, our simulations are not strictly self-similar because of the presence of the axis and its spine, so we will also explore $\alpha$ slightly smaller than the fiducial BP82 value $\alpha=12/16$. As will be discussed later on, values of $\alpha \geq 1$ are numerically problematic. The mass load parameter $\kappa$ is varied between $0.05$ and $1$, which is a range globally consistent with BP82 and F97 jets, both solutions leading to a flow recollimation toward the axis.

The MHD equations have been solved with PLUTO and the results will be presented in dimensionless units. Unless otherwise specified, lengths are given in units of $R_d$, velocities in units of $V_{Kd}= \sqrt{GM/R_d}$, time in units of $T_{d}= R_d/V_{Kd}$, densities in units of $\rho_a$, magnetic field in units of $B_d= V_{Kd}\sqrt{\mu_o \rho_a}$, mass flux in units of $\dot M_d= \rho_a R_d^2 V_{Kd}$ and power in units of $P_d = \rho_a R_d^2 V_{Kd}^3$. In order to be more specific, we translate these quantities for the case of a young star, assuming a one solar mass star with an innermost disk radius $R_d= 0.1$ au, namely 
\begin{eqnarray}
        V_{Kd} & = &94.3 \left( \frac{M}{M_\odot} \right)^{1/2} \left( \frac{R_d}{0.1 au} \right)^{-1/2} \, \mbox{km.s}^{-1} \nonumber \\  
        \dot M_d & = &3.3\, 10^{-10} \left( \frac{\rho_a}{10^{-15} g.cm^{-3}} \right) \left( \frac{M}{M_\odot} \right)^{1/2} \left( \frac{R_d}{0.1 au} \right)^{3/2} \, M_\odot.\mbox{yr}^{-1} \nonumber \\
        P_{jet} & = &6.7\, 10^{41} \left( \frac{\rho_a}{10^{-15} g.cm^{-3}} \right) \left( \frac{M}{M_\odot} \right)^{3/2} \left( \frac{R_d}{0.1 au} \right)^{1/2} \, W \nonumber \\
        B_d & = &10.6 \left( \frac{\rho_a}{10^{-15} g.cm^{-3}} \right)^{1/2} \left( \frac{M}{M_\odot} \right)^{1/2} \left( \frac{R_d}{0.1 au} \right)^{-1/2} \, \mbox{G} \nonumber \\
            T_d & = &1.8 \left( \frac{M}{M_\odot} \right)^{-1/2} \left( \frac{R_d}{0.1 au} \right)^{3/2} \, \mbox{days} 
\end{eqnarray}

The list of all the simulations done in this paper is provided in Table~\ref{tab:ParametresSimus}, with their input parameters $\alpha$ and $\kappa$ and several quantities that are measured at the final stage $t_{end}$ of the simulation. As explained in section 2.4.1, the values of $\mu$ and $\delta$ are dictated by the values of $\kappa$. As will be discussed below, all our simulations display several recollimation shocks. In the table, we provide only the altitude (measured at the axis) of the first main recollimation shock $Z_{shock}$. Since stationary jets require them to become super-Afv\'enic and super-fast magnetosonic (hereafter super-A and super-FM, respectively), we also display the colatitudes $\theta^{ext}_A$ and $\theta^{ext}_{FM}$ of the intersection of the outer boundary $R_{ext}$ and respectively the Alfv\'en and FM surfaces. The last super-FM magnetic surface (defining the jet) can then be followed down to the disk, allowing us to identify the largest anchoring radius $r_{o,FM}$ that we will consider in the JED. This allows us to measure the mass flux emitted from the JED as $\dot M_{jet}= \int_{R_{d}}^{r_{o,FM}} \rho {\bf u}\cdot {\bf dS}$ and compare it with the mass loss emitted from the spine only $\dot M_{spine}= \int_0^{\pi/2} \rho {\bf u}\cdot {\bf dS}$. We also compute the power emanating from the jet $P_{jet}= \int_{R_{d}}^{r_{o,FM}} \rho E {\bf u}\cdot {\bf dS}$ and compare it to the power emanating from the spine $P_{spine}= \int_0^{\pi/2} \rho E {\bf u}\cdot {\bf dS}$.
  
Simulations K1 to A5 were done with a non rotating spine, namely $\Omega_a=0$ and $e_a=2$. Our reference simulation K2 is extensively analyzed in the next section. This reference was repeated with a lower resolution in K2l, all other things being equal, to verify numerical convergence.  Section ~\ref{sec:param} addresses the influence of $\kappa$ (simulations K1 to K5) and $\alpha$ (simulations A1 to A5). In Sect.~\ref{sec:spine} the effect of a rotating spine (simulation SP) will be shortly addressed.

\section{The Blandford \& Payne case}
\label{sec:ref}

\begin{figure*}
\centering
    \includegraphics[trim=0 10 0 17,clip,width=.52\linewidth]{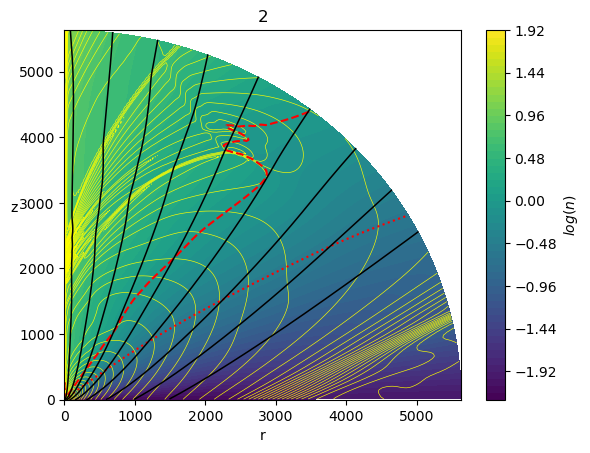}
    \includegraphics[trim=0 10 0 19,clip,width=.45\linewidth]{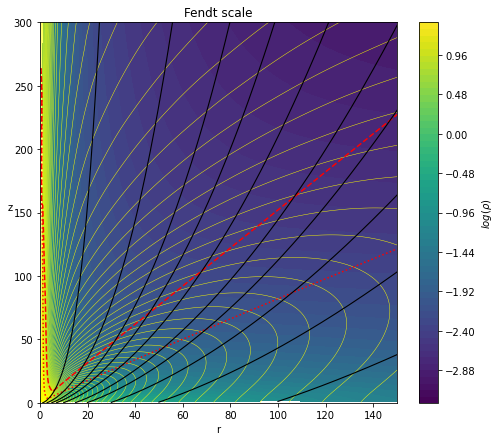}
\caption{Snapshot at $t_{end}$ of our Blandford \& Payne simulation K2. Left: global view with field lines anchored on the disk at $r_o= 3; 15; 40; 80; 160; 320; 600; 1000; 1500$, where the background is the logarithm of the FM mach number $n$. Right: close-up view of the innermost regions, where the background is the logarithm of the density. In both panels, black solid lines are the poloidal magnetic surfaces, the yellow solid lines are isocontours of the poloidal electrical current and the red dashed (resp. dotted) line is the FM (resp. Alfv\'en) critical surface. } 
\label{fig:ReferenceSimulation}
\end{figure*}

\begin{figure}
    \centering
    \includegraphics[width=\linewidth]{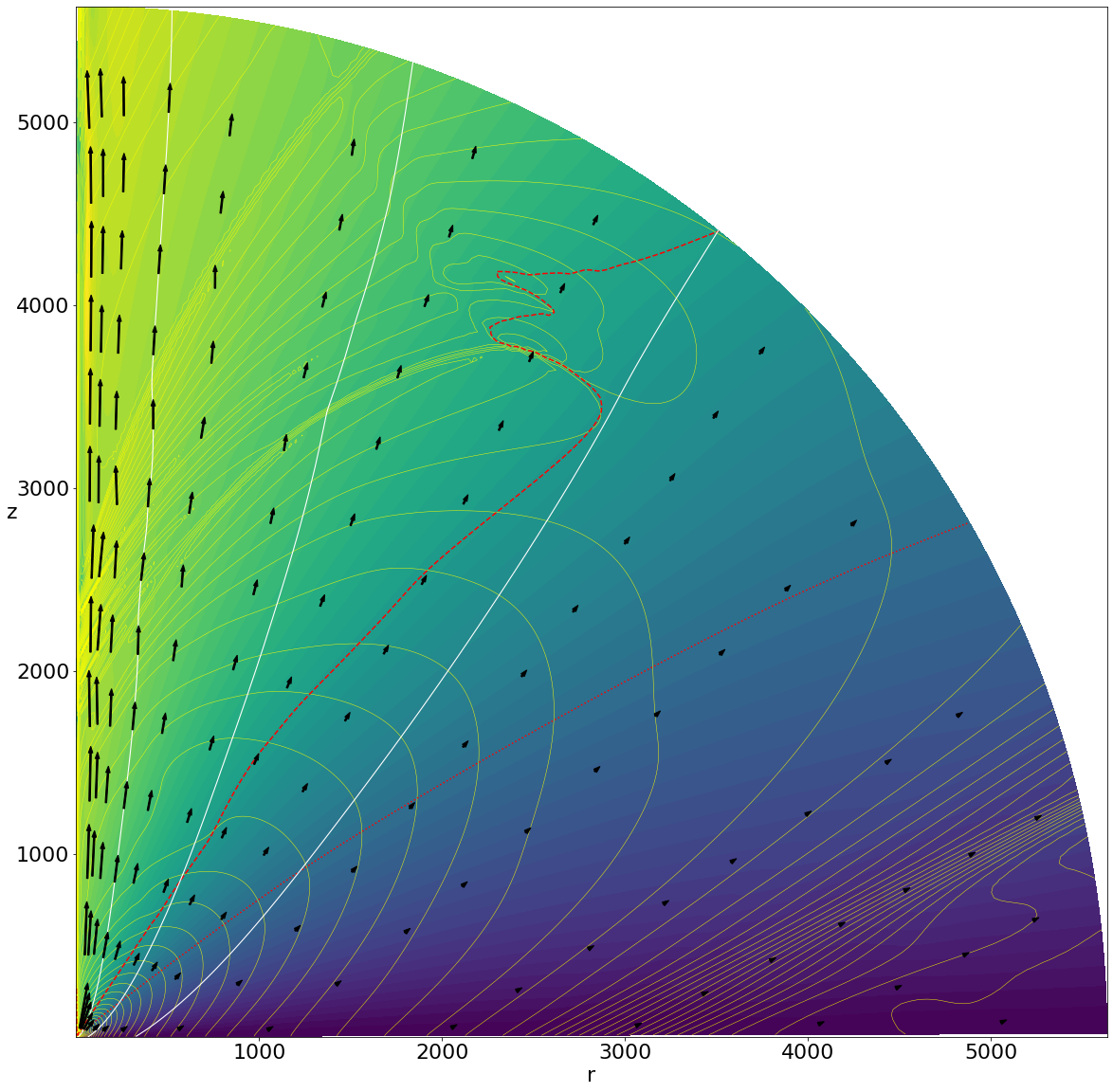}
    \caption{Snapshot of our reference simulation K2 at $t_{end}$. We use the same color coding as in Fig.2, left. The black arrows show the poloidal velocity. The white lines are streamlines inside which (from left to right) 50\%, 75\% and 100\% of the super-FM (spine+jet) mass outflow rate is carried in.}  
    \label{fig:VelocityVectors}
\end{figure}

\subsection{Overview} 

In this section we discuss our reference simulation K2, done for the BP82 $\alpha=3/4$ magnetic field distribution and a mass loading parameter $\kappa=0.1$. It was run up to $t_{end}= 6.5\, 10^5\, T_d$ and has reached a steady-state in a sizable fraction of our computational domain (a quarter of an orbit has been done at $r_o=R_{ext}$). 

Figure~\ref{fig:ReferenceSimulation} displays the final stage reached by K2 at $t_{end}$. The black solid lines are the poloidal field lines, the dotted red line is the Alfv\'en surface (where the Alfv\'enic Mach number $m= u_p/V_{Ap}$ is equal to unity) and the dashed red line is the FM surface (where the FM Mach number $n= u_p/V_{fm}=1$). The left panel shows our simulation on the full computational domain, with a close-up view on the scale used by \citep{fendt2006} on the right panel. The background color is the logarithm of $n$ on the left and the logarithm of the density on the right. The last magnetic surface characterizing the super-FM jet is anchored at $r_{o,FM}=323$ in the JED, and the critical surfaces (A and FM) both achieve a conical shape over a sizable fraction of the domain, which is characteristic of a self-similar steady-state situation. Our spine achieves also super-FM speed at an altitude $z\sim 260$.

The poloidal velocity vectors can be seen on Figure~\ref{fig:VelocityVectors}. The velocity field decreases radially very rapidly mirroring the injection conditions, going from $3.5$ at the spine to $0.2$ at the edge of the super-FM zone (in $V_{Kd}$ units). The white lines show streamlines inside which 50\%, 75\% and 100\% (from left to right) of the total super-FM mass outflow (spine + jet) rate is being carried in. These lines are respectively anchored in the disk at $r_0=10$, $r_o=66$ and $r_{o,FM}=323$. Since $d \dot{M}/dR =  2 \pi R V_z$ falls off very rapidly, this plot shows that even ejection from a very large radial domain may be observationally dominated by the innermost, highly collimated regions up to $r_o\sim 10$, the outer "wide angle wind" remaining probably barely detectable.


The yellow solid lines are isocontours of the poloidal electric current $I= 2\pi r B_\phi/\mu_o$. These contours are very useful as they allow to grasp several important features of the simulation: (1) the typical butterfly shape of the initial accelerating closed electric circuit can be seen up to a spherical radius $R\sim 3000$; (2) for disk radii $r_o \ga 2000$, the electric current flowing out of the disk reaches the outer boundary (most of it in the sub-A regime at high colatitudes) and re-enters in the jet at smaller colatitudes, in the super-FM regime; (3) more importantly, several current sheets can be clearly seen (as an accumulation of current lines), highlighting the existence of several standing (stationary) recollimation shocks. To our knowledge, this is the first time that simulations of super-FM jets exhibit such patterns, predicted in analytical jet studies. Justifications for this assessment are provided section 5.2.      

\begin{figure}
\centering
  \includegraphics[width=.95\linewidth]{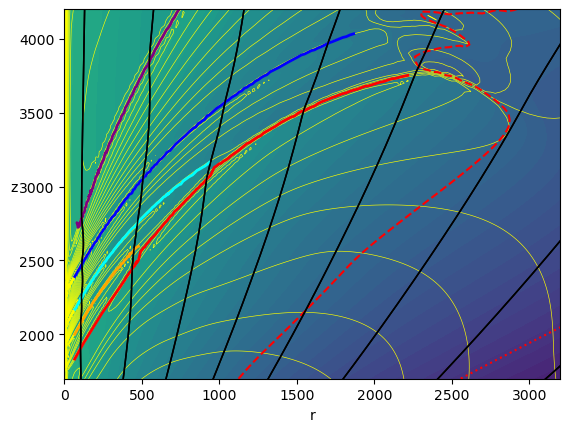}
\caption{Close-up view of our reference simulation K2 at $t_{end}$ showing the shock forming region, with field lines anchored at $r_o= 1.2; 2; 3; 4; 5; 7; 9$. The five shocks are highlighted in red, orange, cyan, blue and purple. We use the same color coding as in Fig.~\ref{fig:ReferenceSimulation}, left.}
\label{fig:ZoomK2}
\end{figure}

These shocks are best seen in Fig.~\ref{fig:ZoomK2}, which presents a zoom around the region of interest. Five shocks (highlighted in colors) can be seen starting near the polar axis, following approximately the expected shape of the MHD characteristics in self-similar jets (see Figs.3 in \citealt{vlahakis2000,ferreira2004}). They are located at $Z_1= 1850, Z_2=2000 , Z_3= 2160, Z_4= 2372$ and $Z_5=2634$.
Only two of these shocks do actually span a significant lateral portion of the jet (those best seen also on the left panel in Fig.~\ref{fig:ReferenceSimulation}). The first one (in red) leaves the axis at an altitude $Z_1=1850$ (labelled $Z_{shock}$ in Tab.~\ref{tab:ParametresSimus}) and stays within our domain, ending up by merging with the FM surface (red dashed curve) around ($r=2500, z=3800$). The second one starts at $Z_5$ and leaves the simulation domain at ($r=1800, z=5200$), being therefore not fully captured by our simulation. For that reason, only the first shock will be extensively described here. 

It can be seen that all shocks do occur only after the magnetic surface has started to bend toward the axis (with a decreasing cylindrical component $B_r$,) and give rise to a sudden outward refraction of the surface with its ouflowing material. This, plus the fact that their positions remain steady in time (see however below), justifies our use of the name "standing recollimation shocks" to describe them. While these standing recollimation shocks appear quite generic for our set of simulations, we stress that they require fantastic spatial and temporal scales to see them.

This simulation was also performed with a resolution two times smaller (see K2l in Table 1). We also observed standing recollimation shocks that are similar, although with less complexity.

\begin{figure}
\centering
  \includegraphics[width=.95\linewidth]{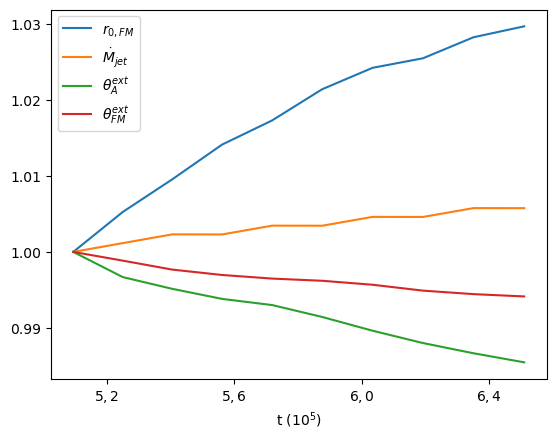}
\caption{Late evolution of several global jet quantities for the simulation K2: the radius $r_{o,FM}$ of the last super-FM surface,  the jet mass loss rate $\dot M_{jet}$ and the two colatitudes $\theta^{ext}_A$ and $\theta^{ext}_{FM}$ that define the position of the two critical surfaces. A slight drift from their initial value is indeed observed. The values provided in Tab.\ref{tab:ParametresSimus} are those achieved at the final time.}
\label{fig:Stationarity}
\end{figure}

\begin{figure}
\centering
    \includegraphics[trim=0 10 0 5,clip,width=.98\linewidth]{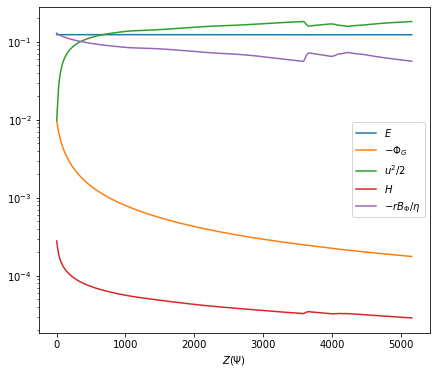}
\caption{Evolution of the various energy contributions along a magnetic surface of anchoring radius $r_{0} = 100$ for the simulation K2 at $t_{end}$:  the Bernoulli invariant $E$, the gravitational potential $\Phi_G$, the total specific kinetic energy $u^{2}/2$, the enthalpy $H$ and the magnetic energy  $-\Omega_{*}rB_{\Phi}/\eta$. The absicssa is the altitude $Z(\Psi)$.}
\label{fig:Energy}
\end{figure}

\subsection{Quasi steady-state jet and spine} 

\begin{figure*}
\centering
\begin{subfigure}{.48\textwidth}
  \includegraphics[trim=0 5 0 5, clip, width=.98\linewidth]{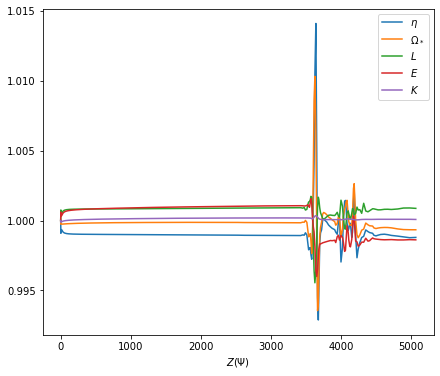}
  \caption{Anchoring radius $r_o=100$.}
\end{subfigure}
\begin{subfigure}{.48\textwidth}
  \includegraphics[trim=0 5 0 5,clip,width=.98\linewidth]{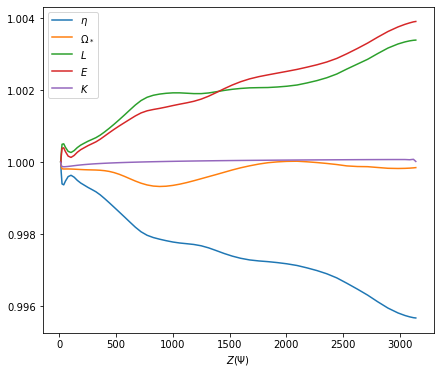}
  \caption{Anchoring radius $r_o=1000$.}
\end{subfigure}
\caption{Evolution of the MHD invariants along field lines of two different anchoring radii $r_{o}$ at $t_{end}$ for simulation K2. All invariants have been normalized to their values at $r_o$. The absicssa is the altitude $Z(\Psi)$.}
\label{fig:Invariants}
\end{figure*}

Jet production is a very rapid process that scales with the local Keplerian time scale (especially since $\mu$ is constant with the radius). It is therefore an inside-out build-up of the jet with its associated electric circuit, until the innermost jet regions (including the spine) reach the outer boundary. This will take a time typically $t_{ext}(r_o=1) \sim R_{ext}/V_z$, with the maximal jet speed $V_z \simeq \sqrt{2\lambda-3} V_K$. According to BP82, cold outflows with moderate inclinations and $\kappa=0.1$ reach $\lambda \sim 10$ (see their Fig.2), which is exactly what we also obtain (see Fig.\ref{fig:LambdaKappa}, with a minimum value $\lambda=11$). This leads to a dynamical time $t_{ext}\sim 1.3\, 10^3$ (in $T_d$ units) so the innermost jet regions achieve an asymptotic state very early. However, the transverse (radial) equilibrium of the outflowing plasma is still slowly adjusting since, as time increases, more and more outer magnetic surfaces achieve their asymptotic state, providing an outer pressure and modifying thereby the global jet transverse equilibrium. This, in turn, necessarily modifies the shape of the magnetic surfaces as well as the associated poloidal electric circuit. 

This means that the MHD invariants along each magnetic surface can always be defined (each surface is quasi-steady), but that they are also slowly evolving in time as the global magnetic structure evolves. For our simulation K2, the last super-FM magnetic surface is anchored at $r_{o,FM}=323$, defining a local Keplerian time $T_K=5.8\, 10^3$. At that distance, the boundary is located at $Z_{ext} \sim R_{ext} \cos{\theta_{FM}} \sim 4430$, leading to a dynamical time $t_{ext}(r_{o,FM}) \sim Z_{ext}/V_z(r_{o,FM}) \sim 1.6\, 10^4$ (with $\lambda=14$) as the speed distribution on the disk is Keplerian. We can therefore expect MHD invariants within our jet to evolve much less only after a time $\sim 10^4$. 

This slight evolution in time of jet quantities is illustrated in Fig.~\ref{fig:Stationarity}. We choose to look at global quantities, such as the radius $r_{o,FM}$ of the last super-FM surface, the jet mass loss rate $\dot M_{jet}$ and the two colatitudes $\theta^{ext}_A$ and $\theta^{ext}_{FM}$ that define the position of the two critical surfaces.  We pick up their values at \textbf{$t= 5.1 10^5$} and plot their evolution by normalizing them to this "initial" value. It can be seen from Fig.~\ref{fig:Stationarity} that their evolution is quite obvious: $r_{o,FM}$ keeps on increasing, leading to an increase in $\dot M_{jet}$ and a decrease of both $\theta^{ext}_A$ and $\theta^{ext}_{FM}$. But the relative variations are less than 3\% for $r_{o,FM}$ and 1\% for the other quantities. 

We therefore consider that our simulation K2 has achieved a fairly global steady-state. In physical units (and for our choice of axial density $\rho_a$) the jet mass loss is about $2\, 10^{-7}\, M_\odot.yr^{-1}$ with a magnetic field around 10 G at 0.1 au. The spine mass loss is only $\sim 10$\% of the jet mass loss, so one can safely presume that the dynamics are mostly controlled by the JED, as expected. However as the spine power is comparable to the jet power ($P_{spine}/P_{jet}=0.81$) the impact of the spine on the collimation and topology of the electric field cannot be neglected. The influence of the spine is detailed in section 4.3.

Figure~\ref{fig:Energy} shows the various contributions to the Bernoulli integral $E(\psi)$ along a magnetic surface anchored at $r_o=100$ at the final stage $t_{end}$. It can be seen that $E$ is indeed conserved and that jet acceleration follows the classical pattern \citep{casse2000}: the kinetic energy (green) increases thanks to the magnetic acceleration, leading to a decrease of the magnetic contribution (magenta). Enthalpy (red) is negligible in this cold outflow. The presence of the shock is clearly seen around \textbf{$Z=3800$}: the flow is suddenly slowed down and the energy is transferred back to the magnetic field, in agreement with the Rankine-Hugoniot jump conditions (see Appendix C). Beyond the shock, MHD acceleration is resumed but, at the edge of our domain, the magnetic field still maintains around 45\% of the initial available energy.
      
The evolution along a magnetic surface of the five MHD invariants is shown in Fig.~\ref{fig:Invariants} for two surfaces, one anchored at $r_o=100$ (left) and the other at $r_o=1000$ (right). In order to plot $\eta, \Omega_*,L, E$ and $S$ on the same figure, we normalize each quantity by its initial (at the disk surface) value at $t_{end}$. On the left, variations of the invariants can indeed be seen but only at the shock located at \textbf{$Z=3800$}. They remain quite tiny, much less than 1\% for all but the entropy (which is conserved to machine accuracy). On the right plot, the field line is anchored beyond $r_{o,FM}$ and the flow remains sub-FM while crossing no shock. Variations of the invariants are again observed, but always less than 0.3\%. This shows that the PLUTO code is quite efficient and the MHD solution is indeed steady.


\begin{figure}
\centering
  \includegraphics[width=.95\linewidth]{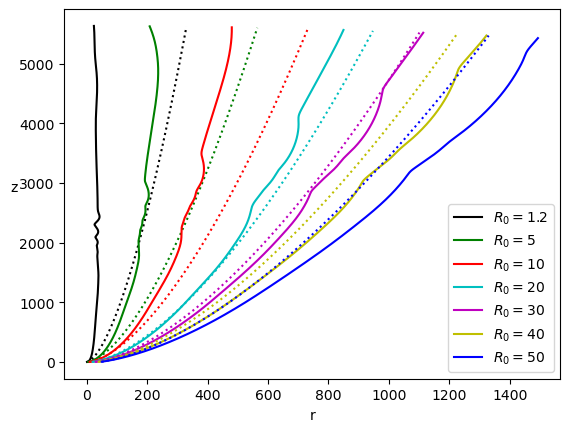}
\caption{Evolution of several magnetic field lines during the simulation computation, for different anchoring radii $R_0$, for the reference simulation K2. In dotted lines are the field lines at the first output of the simulation (initial conditions). In full lines are the field lines at the last output of the simulation (final state).}
\label{fig:EvolutionFieldLines}
\end{figure}

\subsection{Jet collimation}

Before analyzing the shocks in the next section, let us have a word on jet collimation. Figure~\ref{fig:EvolutionFieldLines} shows the initial magnetic field configuration (in dotted lines) along with the final one (solid lines), obtained at $t_{end}$. Each color is associated with a different anchoring radius $r_o$, allowing us to see the evolution from the initial potential field and the final full MHD solution. This plot clearly illustrates how magnetic collimation works. Since the poloidal electric circuit responsible for the collimation must be closed, its sense of circulation must change within the whole outflow (defined as both the jet and its spine). The poloidal current density ${\bf J}_p$ is therefore downward in the inner regions and outward in the outer jet regions. As a consequence, field lines anchored up to $r_o \sim 25$ are focused to the polar axis (Z-pinch due a pole-ward  ${\bf J}_p \times {\bf B}_\phi$ force as $J_z<0$), while field lines anchored beyond $r_o\sim 30$ are de-collimated and pulled out (because of the outward action of the same ${\bf J}_p \times {\bf B}_\phi$ force as $J_z>0$). See Fig.2-right for the topology of ${\bf J}$.

The inner self-collimated jet region can only exist thanks to the existence of these outer, pulled back out jet regions. The final state of the jet collimation, namely the asymptotic jet radius achieved by these inner regions (the densest ones, possibly responsible for the observed astrophysical jets), is therefore also a consequence of the transverse equilibrium achieved by the outflow outskirts with the ambient medium. This balance is mathematically described by the Grad-Shafranov equation and expresses the action of the poloidal electric currents, how they are flowing and how electric circuits are closed within the jet, on the shape of the magnetic surfaces (\citealt{heyvaerts1989}, F97, \citealt{okamoto2001, heyvaerts2003,heyvaerts2003a}). We will come back to this point later on.

\subsection{Standing recollimation shocks} 

\begin{figure*}
\centering
\begin{subfigure}{.48\textwidth}
  \includegraphics[width=\linewidth]{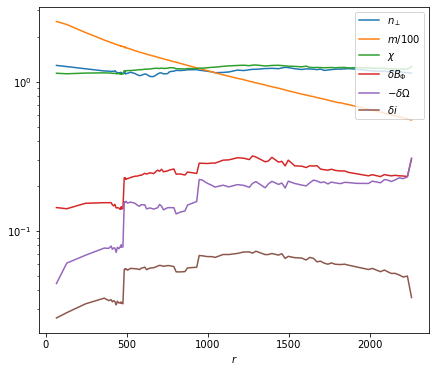}
\end{subfigure}
\begin{subfigure}{.48\textwidth}
  \includegraphics[width=.98\linewidth]{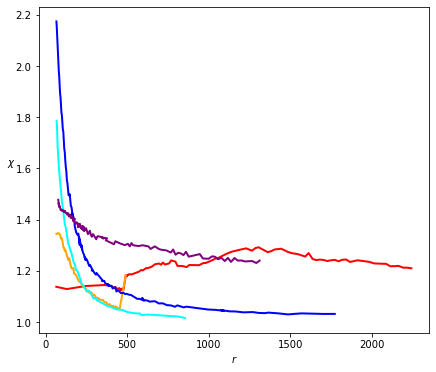}
\end{subfigure}
\caption{Distributions along the shock of several quantities for K2 at $t_{end}$.   
Left: normal incident FM Mach number $n_\perp$,  Alfv\'enic Mach number $m$, compression rate $\chi$, relative variations of the toroidal magnetic field  $\delta {B_{\phi}}$ and plasma angular velocity $- \delta {\Omega}$, and total deviation $\delta i$ (in rad) of the poloidal magnetic field line for the main recollimation shock.  
Right: compression rates $\chi$ of all shocks appearing in Fig.~\ref{fig:ZoomK2}, using the same color code. The main shock corresponds to the red curve. See Appendix C for more details on this figure. }
\label{fig:ReferenceSimulationMainShockRH}
\end{figure*}

Our reference simulation K2 ends up with 5 standing shocks, of which only the first (red in Fig.~\ref{fig:ZoomK2}), starting on the axis around $Z_1= 1850$, will be hereafter studied. This is for two reasons: \\
(1) contrary to the orange and cyan shocks, the red shock surface covers a large extension in the jet itself. It is therefore most probably related to the dynamics of our quasi self-similar jet, whereas these smaller shocks may be related to the spine-jet interaction.\\  
(2) it remains far away from the outer boundary (it ends up at the FM surface at a point $r=2500, z=3800$), which is clearly not the case for the purple shock and possibly also for the blue one.

\subsubsection{Rankine-Hugoniot conditions} 
All shocks are due to the flow heading toward the axis (with decreasing, usually negative $B_r$ and $u_r$ components) at a super-FM velocity, resulting in a sudden jump in all flow quantities with an outward refraction of the magnetic surface (this can be seen in Fig.~\ref{fig:EvolutionFieldLines}). The Rankine-Hugoniot jump conditions (see Appendix C for more details) are of course satisfied with the shock-capturing scheme of PLUTO and MHD invariants are conserved (up to some accuracy as discussed previously). 

The left panel of Fig.~\ref{fig:ReferenceSimulationMainShockRH} displays the evolution of several quantities along the red shock surface. It can be seen that the compression rate $\chi$ (green curve), defined as the ratio of the post-shock to the pre-shock densities, is larger than unity while remaining small ($\leq1.3$, see red curve in the right panel), despite a very large Alfv\'enic Mach number $m \sim 100$ (orange). This is probably because the shock is oblique and the incoming jet material does not reach very large super-FM Mach numbers $n_\perp$  (defined as the ratio of the flow velocity normal to the shock surface to the FM phase speed in that same direction, blue curve). In fact, $n_\perp \leq n= u_p/V_{fm}$ and at these distances, the jet has reached its asymptotic state with a maximum velocity of $u_p \sim \sqrt{2\lambda -3} V_{Ko}$ ($V_{Ko}= \Omega_* r_o$ the Keplerian speed at the anchoring radius). Assuming $B_p << |B_\phi|$, $m>>1$ and a jet widening such that $r>>r_A$ where $r_A$ is the Alfv\'en cylindrical radius along a flow line where $m=1$, leads to 
\begin{equation}   
 n \simeq \frac{u_p}{V_{A\phi}} = \frac{u_p}{\Omega_*r} \frac{m}{1- r_A^2/r^2} \simeq m \frac{r_o}{r} \sqrt{2\lambda-3}\sim m\frac{r_A}{r}   
\end{equation}
which shows that the asymptotic FM Mach number critically depends on how much the jet widens (see \citealt{pelletier1992} and Sect.~ 5 in F97). In our case $n \sim 4$ at the outer edge of the spine-jet interface, in agreement with self-similar studies. Following the main red shock along growing $r$, the incident angle\footnote{Here and in all the following, the angles of incidence and refraction are defined as in Snell-Descartes law, e.g. measured from the normal to the shock front.} of the magnetic field lines on the shock front decreases until turning into a normal shock on its external edge ($r \sim 2000$). Hence, on this edge the shock front coincides with the FM critical surface $n = 1$. As the shock becomes normal, $n_{\perp} \rightarrow n=1$ and the shock vanishes, with a compression rate $\chi$ going to 1.

The three other curves in the left panel of Fig.~\ref{fig:ReferenceSimulationMainShockRH} describe other modifications in jet dynamics. The brown curve is the magnetic field line deviation at the shock front, $\delta i=i_2-i_1$, where $i$ is the flow incidence angle to the normal to the shock surface (subscripts 1 and 2 refer to the pre- and post-shock zones, respectively). The maximum deviation of $0.07 rad=4^\circ$ is very small, in agreement with the small compression rate. 

The purple curve describes the relative variation of the flow rotation $\delta {\Omega}=(\Omega_2-\Omega_1)/\Omega_1$, which is always negative. The shock introduces a sudden brake in the azimuthal speed, so that the compressed shocked material is always rotating less. Since the detection of rotation signatures in YSO jets is an important tool allowing to retrieve fundamental jet properties (see e.g. \citealt{anderson2003,ferreira2006,louvet2018,tabone2020}), recollimation shocks appear to be a very interesting means to lower the jet apparent rotation. However, the rather weak shock found here only introduces a decrease of about $\sim 20 \%$ at the outer edge of the shock. 

The plasma loss of its angular momentum at the shock is of course compensated for by a gain of magnetic field (the angular moment is a MHD invariant). This means that the magnetic field lines are more twisted after than before the shock, as illustrated in the red curve showing  $\delta {B_{\phi}}=(B_{\phi_2}-B_{\phi_1})/B_{\phi_1} >0$. The shock surface acts therefore as a current sheet with an electric current density flowing outwardly (in the spherical $R$ direction).

\subsubsection{Two families of shocks} 
The right panel of Fig.~\ref{fig:ReferenceSimulationMainShockRH} displays the compression rate $\chi$ for the five shocks seen in Fig.~\ref{fig:ReferenceSimulation}, using the same color code. All shocks but the red one have larger compression factors near the axis. The orange and cyan shocks merge with the main red one (respectively at $r\sim 500$ and $r\sim 900$), leading to an increase of its compression rate $\chi$. The large blue shock has the same behavior as the orange and cyan but remains alone (ie not merging with the red) with $\chi$ converging to 1, while the purple shock seems to have a behavior similar to the red one, maintaining a larger value for $\chi$. Although these last two shocks are probably affected by their proximity with the outer boundary, it seems that two classes of recollimation shocks are at stake. 

The first class (represented by the red and purple shocks) corresponds to the recollimation shock predicted in self-similar studies (FP97,\citealt{polko2010}). The reason for their existence is the hoop-stress that becomes dominant as the jet widens, leading to a magnetic focusing toward the axis. As shown in FP97, such a situation always arises in the super-FM regime so that a shock is the only possibility for the converging flow to bounce away. However, as long as no dissipation is introduced, such a situation will repeat. Indeed, after the flow refraction due to the shock, the magnetic field starts to accelerate the plasma again, the magnetic surface widens and the same situation repeats. One would therefore expect periodic oscillations and shocks on a vertical scale $H_R$ (measured on the axis). Figure~\ref{fig:EvolutionFieldLines} provides some evidence of this pattern for the field lines anchored at $r_o=20, 30$ or $40$. The first recollimation shock (red) is quite far away from the disk \textbf{($Z_{shock}=1850$)}, but the second shock (purple) occurs at $Z_5\simeq 2634$. A much larger computational domain would be necessary in order to clearly assess a periodic pattern.       
     
\begin{figure}
\centering
\includegraphics[width=.98\linewidth]{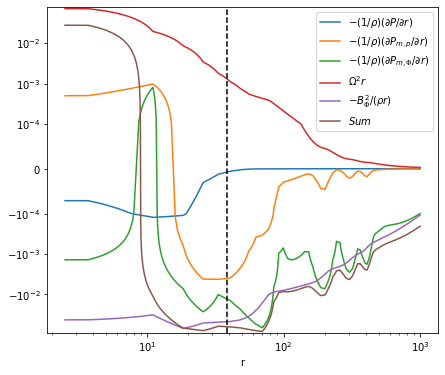}
\caption{Radial distribution of the radial accelerations and their sum at the altitude $Z=2400$ for the simulation K2 at $t_{end}$. The vertical dashed line corresponds to the spine-jet interface, namely the field line anchored at $r_o=R_d$.}
\label{fig:RadialForces}
\end{figure}

The second class of recollimation shocks (represented by the orange, cyan and blue shocks) are limited to the vicinity of the spine-jet interface and are thereby a consequence of a radial equilibrium mismatch between these two super-FM outflows. The transverse equilibrium of a magnetic surface is provided by projecting the stationary momentum equation in the direction perpendicular to that surface, leading to the equation (FP97)
\begin{equation}
    (1-m^2) \frac{B_p^2}{\mu_0 \mathcal{R}} - \nabla_\perp \bigg( P+ \frac{B^2}{2 \mu_0} \bigg) - \rho \nabla_\perp \Phi_G + \bigg(\rho \Omega^2 r - \frac{B_\phi^2}{\mu_0 r} \bigg) \nabla_\perp r = 0 \ .
\end{equation} 
Here, $\nabla_\perp= {\bf e}_\perp \cdot \nabla$ provides the gradient perpendicular to a magnetic surface with ${\bf e}_\perp= \nabla \Psi/|\nabla \Psi|$ and $B_p^2/\mathcal{R} = {\bf e}_\perp \cdot ({\bf B}_p\cdot\nabla {\bf B}_p)$, measures the local curvature radius $\mathcal{R} $ of the magnetic surface. In the asymptotic region where these small recollimation shocks are observed, the field lines are almost vertical and gravity is negligible. The above equation reduces then to    
\begin{equation}
     - \frac{\partial}{\partial r} \bigg( P+ \frac{B^2}{2 \mu_0} \bigg) + \rho \Omega^2 r - \frac{B_\phi^2}{\mu_0 r}  = 0 \ .
\label{eq:transfield}
\end{equation}
Looking at Fig.~\ref{fig:RadialForces}, where the various forces are plotted as function of the cylindrical radius at a constant height $Z=2400$, it can be clearly seen that the dominant force is the hoop-stress $- B_\phi^2/(\mu_or)$ (purple curve) near the spine-jet interface (dashed vertical line), defined as the magnetic field line anchored at $r_o=R_d$. That pinching force overcomes the others (the above sum is actually non-zero and negative), indicating that the field lines do have a curvature and are actually converging toward the axis (thus Eq.~\ref{eq:transfield} above is too simplistic). Nevertheless, this behavior of the forces is consistent with the stationary shape of the magnetic surfaces seen at $Z=2400$ at those radii. Further out downstream, the fifth (purple) shock will make the field lines bounce back again. This tells us that we are witnessing radial oscillations of the radius of the spine driven by a mismatch between the dominant forces (hoop-stress, magnetic pressure and centrifugal term). 

This oscillatory behavior may be a generic feature of MHD outflows from a central rotator as shown by \citet{vlahakis1997}, because of the different scaling with the radius of the pinching force and the centrifugal force (as in the self-similar jet, F97). But it may also be triggered by the pinching due to the outer jet recollimation, namely the spine-jet interface response to the global jet recollimation. Indeed, no spine-jet shock is seen before the onset of the main recollimation shock. We further note that the five shocks are located at a slightly increasing distance from each other. Indeed, $\Delta Z_{12}=  Z_2-Z_1= 150, \Delta Z_{23}= 160, \Delta Z_{34}= 212$ and $\Delta Z_{45}= 262$, which is the sign of some damping of the spatial oscillations at the spine-jet interface (see e.g. \citealt{vlahakis1997}).  This corresponds to three spine-jet shocks (orange, cyan and blue) located between the two large jet recollimation shocks (red and purple), as can be seen in Fig.~\ref{fig:ZoomK2}.   

Despite the fact that the magnetic surfaces are in a steady state, it is useful to look at this spatial oscillatory pattern as the non-linear outcome of transverse waves. Right after a shock, the flow is again outwardly accelerated leading to a widening of the magnetic surface and its refocusing toward the axis with the unavoidable shock. It will thus take a time $\Delta t_z= \Delta {Z}/u_z$ to reach the next shock. On the other hand, any radial unbalance triggered right after the shock gives rise to a transverse (radial) FM wave that bounces back on a time $\Delta t_r= 2 r/V_{fm}$ measured at the spine-jet interface. In steady-state, these two times must be the same, which requires $\Delta {Z} \sim 2n r$, where $n$ is the FM Mach number measured at the spine-jet interface. At $Z=2400$, the width of the spine is $r \simeq 40$ and $n \simeq 3$ providing the correct order of magnitude for $ \Delta {Z}\sim 240$ . This is also verified for all other shocks. This correspondance strengthens the idea that these small shocks are actually triggered by the first large recollimation shock.
   

\subsection{Electric circuits} 

\begin{figure}
\centering
\includegraphics[width=.98\linewidth]{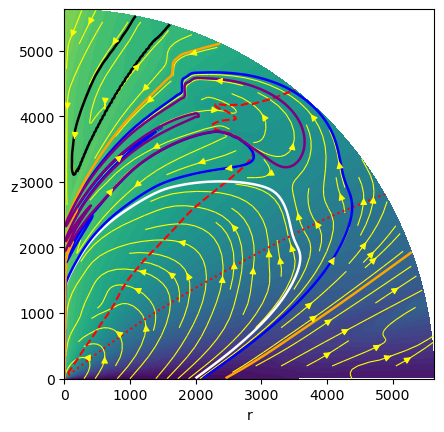}
\caption{Plot of the poloidal electric circuits at $t_{end}$ for simulation K2. The two red curves are the critical surfaces, Alfv\'en (dotted) and FM (dashed). The yellow curves are the poloidal electric circuits, defined as isocontours of $r B_{\phi}$, the arrows indicating the direction of the poloidal current density ${\bf J}_p$. Four circuits are especially highlighted: (1) the envelope of the inner accelerating current in white ($r B_{\phi}=-2.06$), (2) the outermost circuit still fully enclosed within the domain in blue ($ B_{\phi}=-2.005$), (3) a circuit closed outside the domain in orange ($r B_{\phi}=-1.80$) and (4) a post-shock accelerating circuit in purple (also with $r B_{\phi}=-2.06$).}  
\label{fig:ReferenceSimuArrows}
\end{figure}

The existence of these shocks drastically affects the poloidal electric circuits that go along with MHD acceleration (and of course collimation). This can be seen in Fig.~\ref{fig:ReferenceSimuArrows}, where several interesting circuits are evidenced in color. Each poloidal circuit corresponds to an isocontour of  $r B_\phi$, the arrows indicating their flowing direction.

The white contour marks the last electric circuit that flows below the first recollimation shock and defines the envelope of the initial accelerating current. It links the disk emf with the accelerated jet plasma and flows back to the disk along the spine. This is due to the fact that the largest electric potential difference is with the axis.
  
The blue circuit is the last electric circuit fully enclosed within the computational domain. The current flows out of the disk (further away than the previous circuit) and makes a large loop that goes beyond (downstream) the main recollimation shock, flowing back on the axis below $Z_5$ until it encounters the smaller shock at $Z_4$ (blue curve in Fig.~\ref{fig:ZoomK2}). Since a shock behaves as an emf, with an outwardly (positive $J_r$) electric current flowing along its surface, the blue electric current gets around it and goes back to the axis where it meets the next shock surface at $Z_3$ (cyan  in Fig.~\ref{fig:ZoomK2}). Since that shock merges with the main recollimation shock, the blue electric current flows along these two surfaces, gets around the main shock (near the point $r\sim z\sim 3000$ in Fig.~\ref{fig:ReferenceSimuArrows}) and returns back to the disk via the spine, where it joins the white circuit below $Z_1$.  

As said before, the outflowing plasma gets re-accelerated after each shock. This requires a local accelerating electric circuit which is naturally enclosed within two recollimation shocks. One such circuit is exemplified by the purple contour in Fig.~\ref{fig:ReferenceSimuArrows}. It has actually the same $rB_\phi$ value as the white one, but is enclosed between the two large recollimation shocks. Since the small shocks (orange and cyan in Fig.~\ref{fig:ZoomK2}) merge with the main one, the purple accelerating circuit is the envelope of the current used to go from the first main shock to the second one flowing back to the spine just before $Z_5$ (and getting around the shock starting at $Z_4$ near the point $r\sim2000, z\sim 4000$, like the previous blue electric circuit).

These three examples of electric circuits (white, purple and even blue) are fully closed within the computational domain and therefore do not contribute to any further asymptotic collimation. However, it can be seen that the electric current outflowing from the disk beyond $r_o\sim 2000$, leaves the computational domain and is supposedly closed by the inflowing current that enters the computational domain at small colatitudes. One example of such electric circuit is provided by the orange curve in Fig.~\ref{fig:ReferenceSimuArrows}. This inward electric current is responsible for the inner jet collimation at large distances, say at $Z\ga 3000$, and is seen to flow back to the disk along the spine, which acts as a conductor. 
This implies that the asymptotic jet collimation is here somewhat controlled by an electric circuit that is not fully self-consistent. Indeed, this electric circuit is actually determined by the boundary conditions at $R_{ext}$ and there is no guarantee that its evolution is consistent with the disk emf. This is of course unavoidable but it may have an impact on the collimation properties of numerical jets (see discussion in Sec.~\ref{sec:disc}). 
Moreover, since this current embraces very large spatial scales, very long time scales are consistently implied and may lead to an evolution of the jet transverse equilibrium on those scales.

\subsection{Time evolution} 
\label{sec:time}

\begin{figure}
\centering
\includegraphics[width=.98\linewidth]{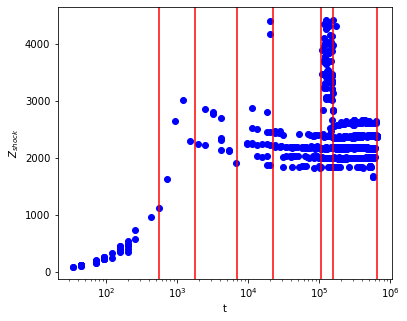}
\caption{Altitude $Z$ of the different shocks (measured at the axis) as function of time (in $T_d$ units) for simulation K2. The vertical lines correspond to the six times $t_i$ used in Fig.~\ref{fig:timeK2}:  $t_1= 551, t_2=2.08\, 10^3, t_3= 8.51\, 10^3, t_4=1.99\, 10^4, t_5=1.05\, 10^5, t_6=1.58\, 10^5$. The last vertical line is $t_{end}=6.51 \, 10^5$.} 
\label{fig:EvolutionShockPosition}
\end{figure}

\begin{figure*}
\centering
\begin{subfigure}{.99\textwidth}
  \includegraphics[trim=0 0 0 80, clip,  height=.35\linewidth]{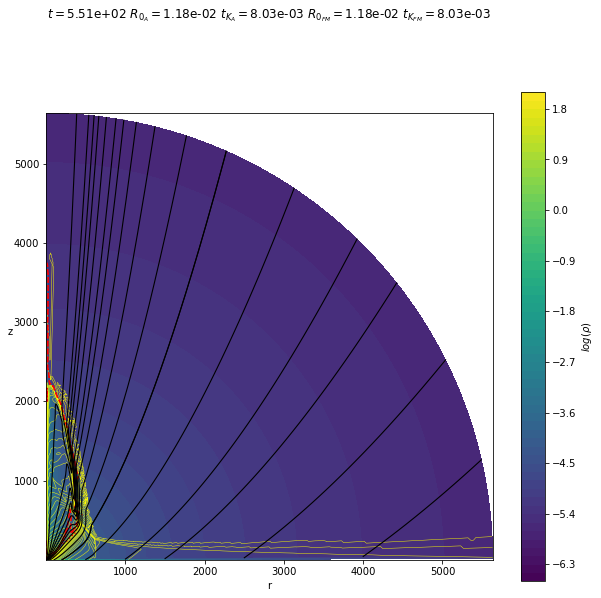} 
  \includegraphics[trim=0 0 0 80, clip, height=.35\linewidth]{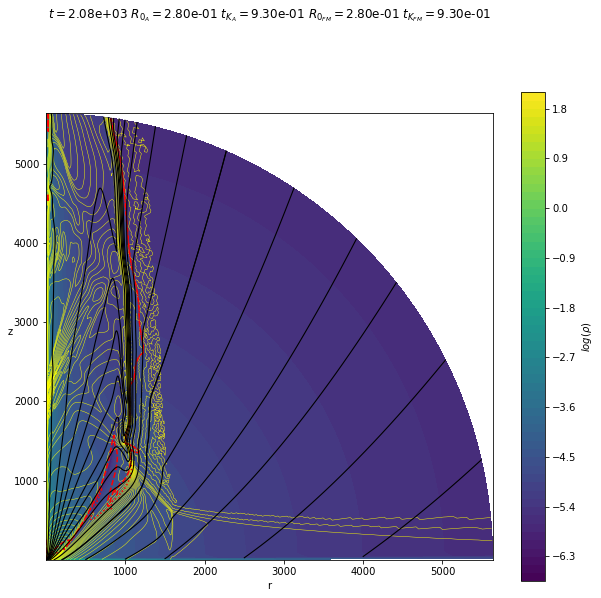} 
\end{subfigure}
\begin{subfigure}{.99\textwidth}
  \includegraphics[trim=0 0 0 80, clip, height=.35\linewidth]{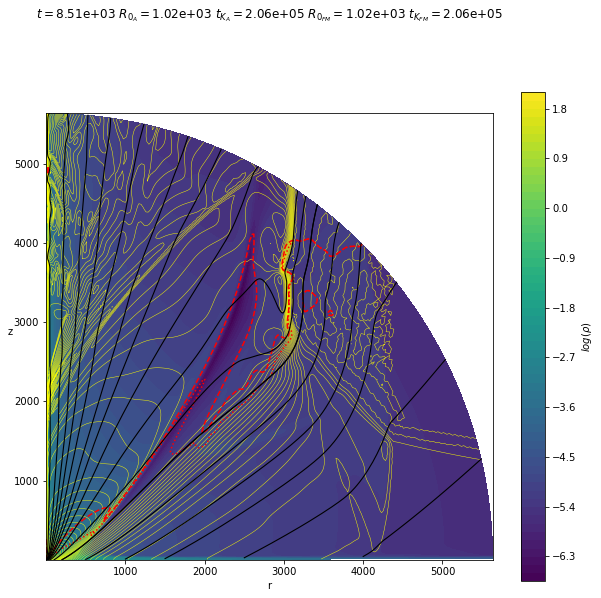} 
  \includegraphics[trim=0 0 0 80, clip, height=.35\linewidth]{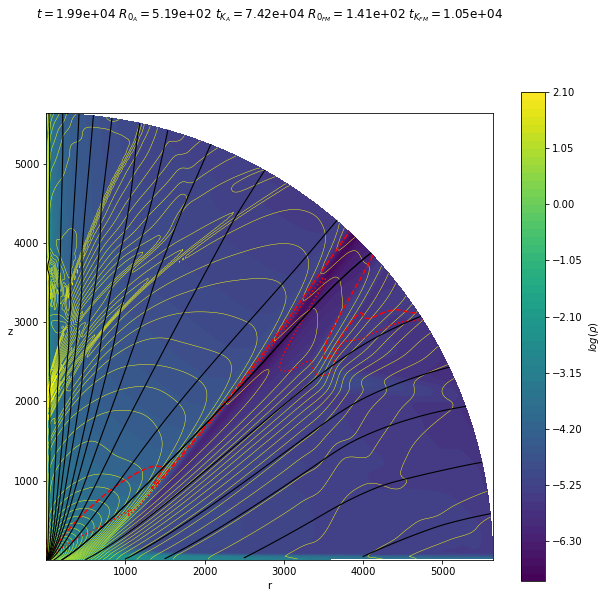} 
 \end{subfigure}
\begin{subfigure}{.99\textwidth}
  \includegraphics[trim=0 0 0 80, clip, height=.35\linewidth]{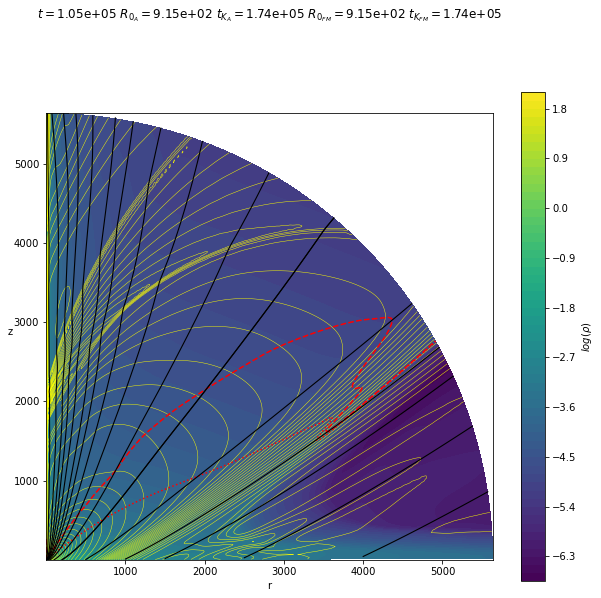} 
  \includegraphics[trim=0 0 0 80, clip, height=.35\linewidth]{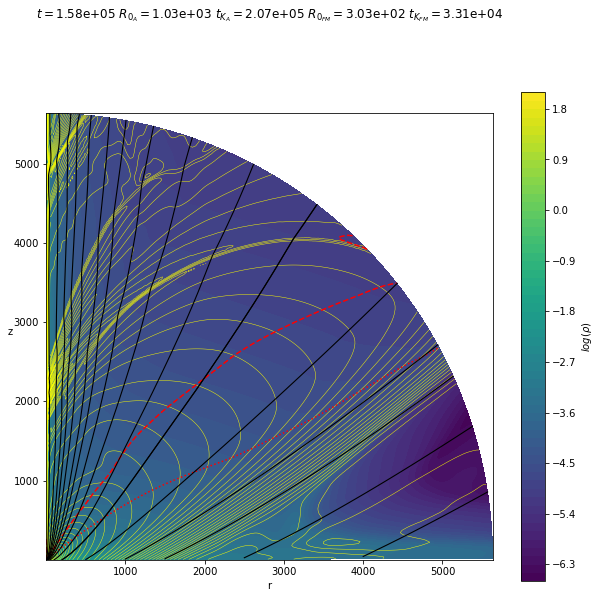} 
\end{subfigure}
\caption{Snapshots of our reference simulation K2 at different times (given in $T_d$ units). From top to bottom, left to right: $t_1= 551, t_2=2.08\, 10^3, t_3= 8.51\, 10^3, t_4=1.99\, 10^4, t_5=1.05\, 10^5, t_6=1.58\, 10^5$. The background color is the logarithm of the density, black lines are the magnetic surfaces, red lines the Alfv\'en (dotted) and FM surfaces (dashed) and yellow curves are isocontours of the poloidal electric current.}
\label{fig:timeK2}
\end{figure*}

Figure~\ref{fig:EvolutionShockPosition} shows the evolution in time of the vertical height $Z$ (measured at the axis) of all shocks found in the simulation. As discussed previously, it takes a time $t_{ext}(1)\sim 10^3$ for the innermost jet (anchored at $r_o=R_d$) to reach the boundary of our computational box. During this early evolution with $t < t_{ext}(1)$, the detected shocks correspond to the first bow shock where the jet front meets the initial unperturbed ambient medium. Once the jet has reached the boundary, the spine is in a steady-state while the jet transverse equilibrium keeps on evolving due to the self-similar increase of its width. Indeed, the time to reach the outer boundary for a magnetic surface anchored at $r_o$ in the disk grows as $t_{ext}(r_o) \propto r_o^{1/2}$. It thus takes a time $\sim 10^4$ to achieve a steady ejection from $r_o=r_{o,FM}= 323$, where $r_{o,FM}$ is the maximum radius of the field lines that achieve a super-FM flow speed. As argued before, steady-state of the global jet structure is only expected beyond that time. 

The vertical lines in Fig~\ref{fig:EvolutionShockPosition} trace six times $t_1= 551, t_2=2.08\, 10^3, t_3= 8.51\, 10^3, t_4=1.99\, 10^4, t_5=1.05\, 10^5$ and $t_6=1.58\, 10^5$. The snapshots corresponding to each of these times are shown in Fig.~\ref{fig:timeK2}, allowing to see the global jet evolution. The times $t_1$ and $t_2$ have been chosen to enclose $t_{ext}(1)$. The bow shock with the ambient medium can be clearly seen, as well as the outward (radial) evolution of the jet width. At $t_2$ several shocks near $Z\sim 2000$ can be seen in both figures. The jet radial equilibrium is clearly not yet steady. However, fig~\ref{fig:EvolutionShockPosition} shows that while the positions of the shocks (as measured on the axis) is already close to their final value, their final number is not yet settled.    

Four standing recollimation shocks seem to settle somewhat between $t_3= 8.51\, 10^3$ and $t_4=1.99\, 10^4$, in agreement with our previous estimate. They can be clearly seen in Fig.~\ref{fig:timeK2}, where some transient shocks located further up at $Z>4000$ at $t_3$ have disappeared at $t_4$. Also, given the huge spatial scales involved, most of our JED is still evolving. For instance, while there has been already $3183$ orbital periods at $R_d$ at $t_4$, the disk has done only half of an orbit at $r_o=323$. This can be seen in the shape of the FM surface, which has not reached yet its steady-state configuration (conical).

Beyond $t_4$, the global flow is slowly evolving in time in some adiabatic way, with four standing recollimation shocks. The evolution of the jet outer regions and the progressive evolution of the A and FM surfaces to their conical shapes seem to produce no obvious evolution in the shocks until $t_5$. At that time, a dramatic evolution is triggered with the appearance of shocks beyond $Z_4$. Figure~\ref{fig:EvolutionShockPosition} clearly shows this pattern with the appearance of a fifth shock. Its altitude $Z_5$ evolves in time, consistently with $Z_4$ until a steady-state is finally reached approximately near $t_6$. Our final state $t_{end}$ shows no relevant difference in the positions of the five shocks. This evolution of the distribution of the shocks has only slightly affected the position of the farthest shock $Z_4$, leading to the final regular distance $\Delta Z$ discussed previously.     

\begin{figure}
\centering
\includegraphics[width=.98\linewidth]{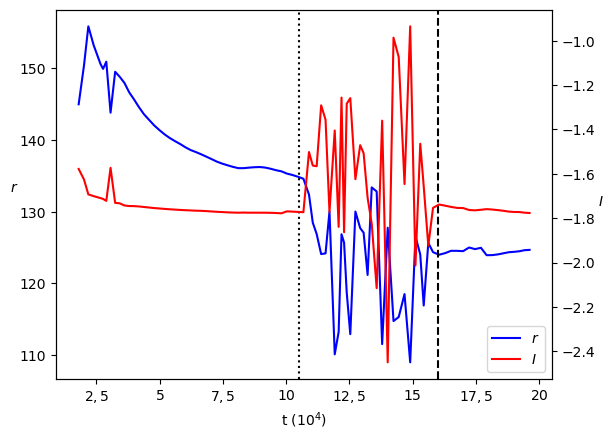}
\caption{Time evolution of the cylindrical radius $r$ measured at $Z=3500$ of the magnetic surface anchored at $r_o=3$ (blue curve) and the electric current $I = r B_{\phi}$ (red) flowing within that surface for simulation K2. The two vertical dashed lines correspond to $t_5=1.05\, 10^5$ and $t_6=1.58\, 10^5$.}
\label{fig:ShockSeparation}
\end{figure}

The appearance of a fifth shock at $t_5$ leading, after a transient phase ending up at $t_6$, to a new steady state jet configuration is illustrated in Fig.~\ref{fig:ShockSeparation}. The time evolution of the cylindrical radius of the field line anchored at $r_o=3$ (in blue) is measured at a constant height $Z=3500$. It can be seen that this radius is steadily slowly decreasing in time, going from $r\sim 150$ near $t_4$ down to $r\sim 135$ at $t_5$ where some fluctuations are suddenly triggered. These oscillations describe a time dependent behavior which ends up at $t_6$, with a new radial balance found at a smaller radius $r\simeq 125$. Globally, this evolution describes a magnetic surface that is first slowly getting more and more confined, then enters an unstable situation until finding out another (tighter) equilibrium. This behavior is consistent with the evolution of the electric current (red curve) that flows within that magnetic surface, which is seen to first steadily (although very slowly) increase until achieving a final value. 

%

In these inner regions, the cold jet transverse equilibrium is mostly achieved by the poloidal magnetic pressure balancing the toroidal pressure and hoop stress (see Eq.~\ref{eq:transfield} and Fig.~\ref{fig:RadialForces}). In this Bennett relation, one gets 
\begin{equation}
 (r B_\phi)^2 = r^2 B_z^2 + \int_0^r 2B_z^2 rdr \ .
\end{equation}  
This relation shows that the toroidal field $B_\phi$ is compelled to follow the same scaling as $B_z$ in order to maintain the jet transverse equilibrium.  If we assume that the self-similar radial scaling for the vertical field $B_z$ is recovered at large distances, we get $(r B_\phi)^2 \propto r^{2(\alpha-1)}$. As a consequence, for $\alpha<1$ (which is the case here), whenever the electric current is increasing, the radius of the magnetic surface decreases (as in a Z-pinch). This scaling provides $\Delta I/I = (\alpha-1) \Delta r/r= -0.25 \Delta r/r$, consistent with the evolution seen in Fig.~\ref{fig:ShockSeparation}.

This slow increase in time of the electric current flowing near the axis is a natural consequence of the increase of the disk emf $e_{disk}$ as the outer disk regions achieve a steady-state. Indeed $e_{disk} \simeq \int_{R_d}^{r_{max}} \Omega_K r B_z dr$ increases with $r_{max}$, which is the maximal radius in the disk that achieved a steady-state. This increase in the available current is expected to stop when no relevant emf is added anymore. The available current is determined at the disk surface by the crossing of the Alfv\'en critical point, since it is that point that fixes the available total specific angular momentum carried away. One can therefore estimate the time where the current should level off as the time when the outermost magnetic field line reached the Alfv\'en point. Figure~\ref{fig:ShockSeparation} shows that this is achieved approximatively around $t_6$ with $r_{max} \sim 10^3$, corresponding to a full orbital period at $r_{max}$ equal to $t = 2\pi r_{max}^{3/2}= 2\, 10^5$ (note that the time for the flow ejected at $r_{max}$ to reach the outer boundary is comparable). After a time of a few $10^5\, T_d$, our simulation has finally achieved a global steady-state.         


\begin{figure}
\centering
\includegraphics[width=.98\linewidth]{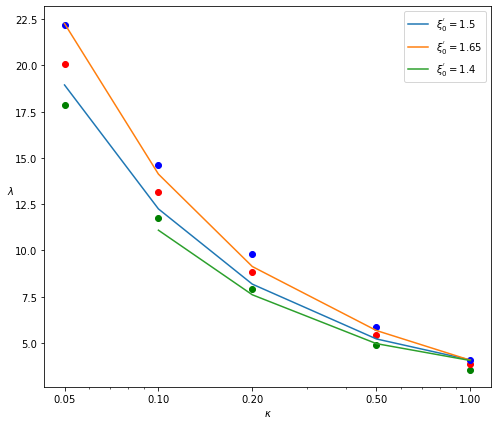}
\caption{Jet parameter space $\lambda(\kappa)$ at the final stage of our simulations K1-K5 with $\alpha=3/4$. Each simulation is obtained for a unique mass loading $\kappa$ and gives rise to a distribution of the magnetic lever arm $\lambda$ with the radius: green, yellow and blue dots correspond to anchoring radii $r_o= 5, 50, 500$ respectively. The solid curves are obtained for constant values (indicated in the panel) of the initial magnetic field inclination $\xi^{'}_{0}=B_r/B_z$ at the disk surface.}
\label{fig:LambdaKappa}
\end{figure}

\section{Parameter dependence}
\label{sec:param}

\begin{figure*}[htp]
\centering
\begin{subfigure}{.8\textwidth}
  \includegraphics[height=.35\linewidth]{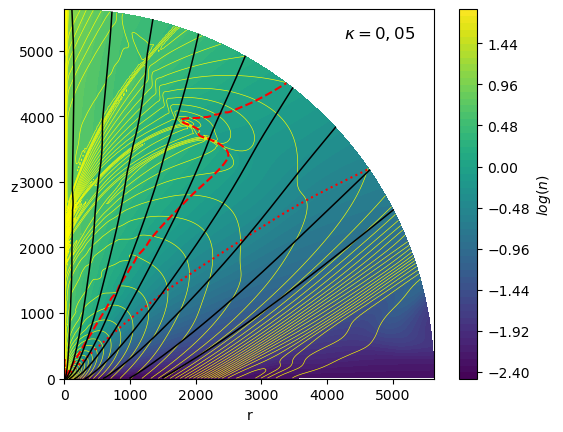}
  \includegraphics[height=.35\linewidth]{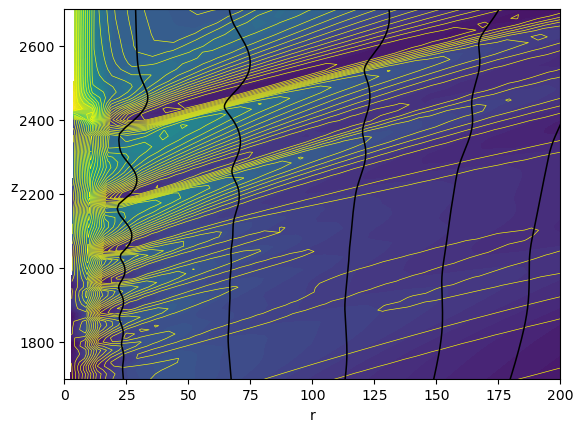}
\end{subfigure}
\begin{subfigure}{.8\textwidth}
  \includegraphics[height=.35\linewidth]{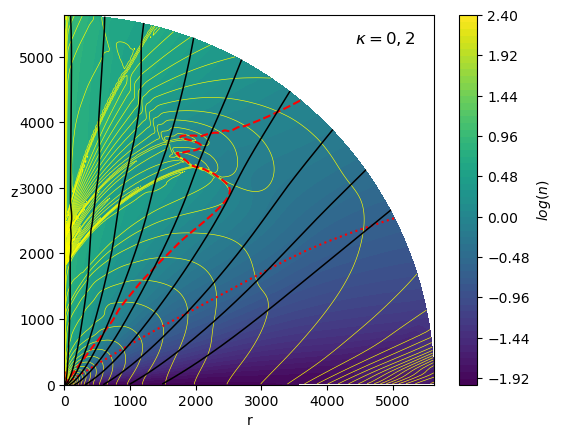}
  \includegraphics[height=.35\linewidth]{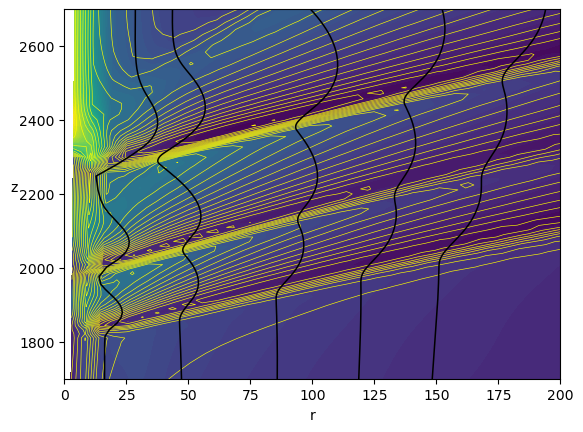}
\end{subfigure}
\begin{subfigure}{.8\textwidth}
  \includegraphics[height=.35\linewidth]{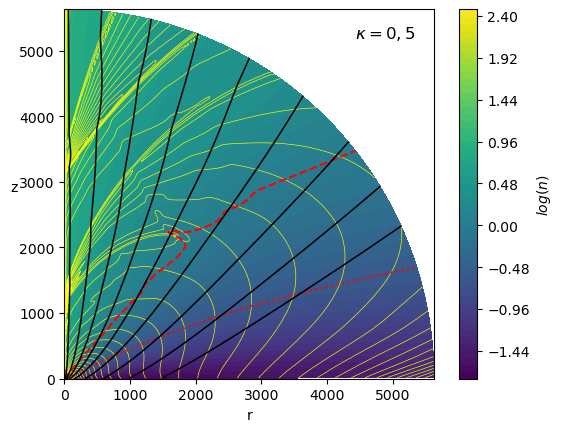}
  \includegraphics[height=.35\linewidth]{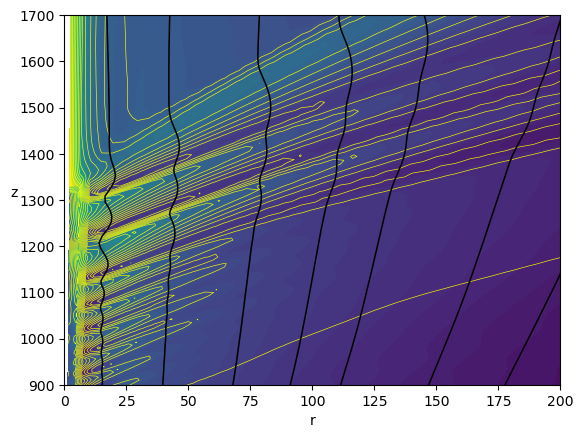}
\end{subfigure}
\begin{subfigure}{.8\textwidth}
  \includegraphics[height=.35\linewidth]{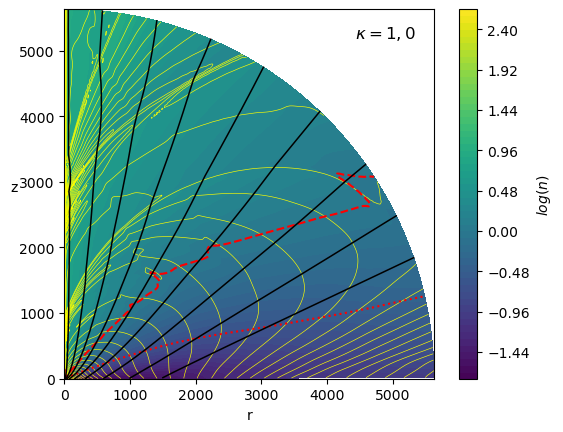}
  \includegraphics[height=.35\linewidth]{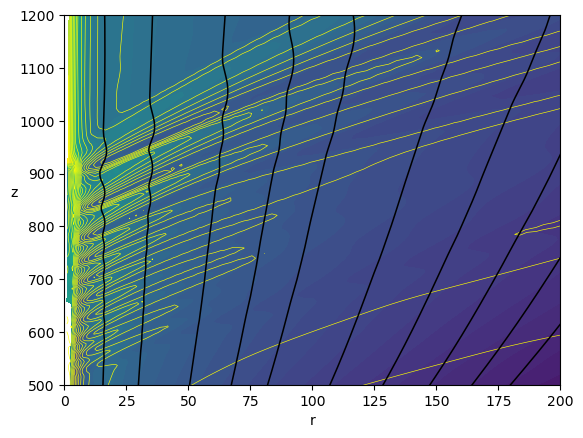}
\end{subfigure}
\caption{Influence of the mass loading parameter $\kappa$ on the final stage of jets obtained with $\alpha=3/4$. The color background is the logarithm of the FM Mach number $n$, black solid lines are field lines, yellow lines are isocontours of the electric current $rB_\phi$ and the red dashed (resp. dotted) curve is the FM (resp. Alfv\'en) critical surface. The left panels show the whole domain and the right panels a close-up view around the shock formation regions. In the left panel, the field lines anchoring radii are  $r_o= 3; 15; 40; 80; 160; 320; 600; 1000; 1500$. In the right panel, the field lines anchoring radii are $r_o= 1.2; 2; 3; 4; 5; 7; 9; 11; 13; 15$.}
\label{fig:SimulationsKappa}
\end{figure*}

In the previous section, it has been shown that our reference simulation K2 behaves qualitatively like the self-similar analytical calculations of \citet{blandford1982} and F97, with a refocusing toward the axis and the formation of a recollimation shock. However, the presence of the axial spine breaks down the self-similarity and introduces additional shocks localized at the spine-jet interface. 
    
To further understand the behavior of these shocks, we conducted a parameter study in $\kappa$ and $\alpha$, respectively the jet mass load and the radial exponent of the disk magnetic flux. We finally did one simulation with the same JED parameters as in our reference simulation K2, but with a rotating spine. All our simulations are presented in Table~\ref{tab:ParametresSimus}.

\subsection{Influence of the mass loading parameter $\kappa$} 

In this section, we present our simulations K1 to K5, obtained with the same parameters as K2 except for the mass load $\kappa$ which is varied from $\kappa=0.05$ to $\kappa=1$. Our parameter range in $\kappa$ is slightly smaller than the one achieved by \citet{blandford1982} which goes down to $\kappa=0.01$. It can be seen in Fig.~\ref{fig:LambdaKappa}, which represents the magnetic lever arm $\lambda$ as function of the mass loading parameter $\kappa$. Each simulation is obtained for a unique value of $\kappa$ but since the simulation is not strictly self-similar, we obtain a range in $\lambda$: the larger the anchoring radius, the larger the field lines inclination at the disk surface and the larger the magnetic lever arm $\lambda$. To ease the comparison with Fig.~2 of \citet{blandford1982}, we computed for each simulation at which anchoring radius $r_o$ the field line inclination $\xi^{'}_{0}=B_r/B_z$ at the disk surface is equal to $1.4, 1.5$ and $1.65$. This allowed us to draw in our Fig.~\ref{fig:LambdaKappa} iso-contours of $\xi^{'}_{0}$, which are in agreement with the above expectations and analytical self-similar calculations (see also Fig.~3 in F97).   

All simulations achieve a steady-state and exhibit basically the same behavior as K2, as can be seen in  Fig. \ref{fig:SimulationsKappa}. From top to bottom, $\kappa$ increases from $0.05$ to 1, the left panels showing the whole computational domain with the two critical surfaces in red (A, dotted and FM dashed) and the right panels providing a close up view around the shock forming regions near the axis. Table~\ref{tab:ParametresSimus}  provides the value of several jet quantities : the spine mass loss rate stays around 10\% of the jet mass loss rate, despite the net increase in $\dot M_{jet}$ ($\propto \kappa$, factor 20 increase). Similarly, while the jet power $P_{jet}$ scales in $\kappa$, the spine power stays around 80\% of the jet power.

\begin{figure}
\centering
\includegraphics[width=.98\linewidth]{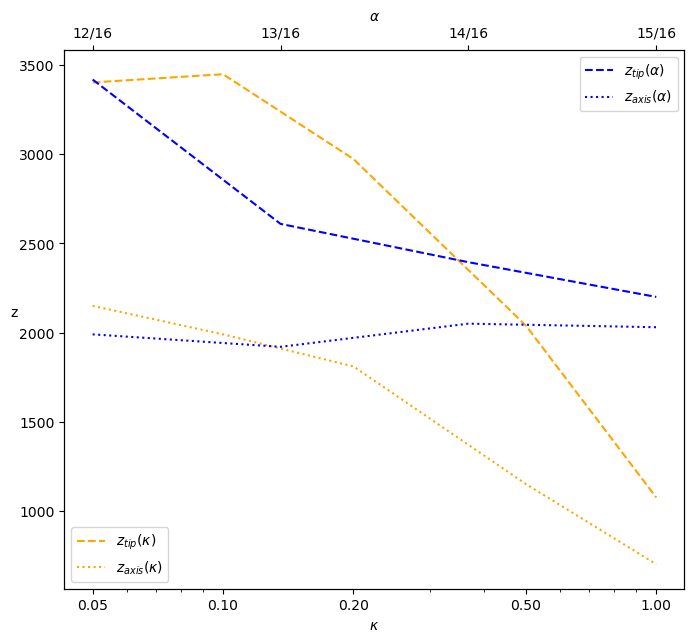}
\caption{Influence of $\kappa$ (orange lines) and $\alpha$ (blue lines) on the altitude of the main recollimation shock. This is done by measuring two altitudes for each shock: its height at the axis ($Z_{shock}$, solid lines) and the altitude of the shock's outer edge ($Z_{tip}$, dashed lines). The scale for $\kappa$ is indicated below, while the scale for $\alpha$ is above.}
\label{fig:KappaShockAltitudeEvolution}
\end{figure}

The first observation is that the altitude of the main recollimation shock (the one merging with the FM surface) decreases globally with $\kappa$. This is quantitatively shown in Fig.~\ref{fig:KappaShockAltitudeEvolution}, where $Z_{shock}$ moves from 2150 down to 700. The same evolution occurs for the altitude $Z_{tip}$ where the main shock merges with the FM surface. Globally, as $\kappa$ increases, the whole jet structure decreases toward the disk. 


Such a behavior is consistent with the self-similar calculations obtained by F97. Indeed, as evidenced in his Fig.~6, the denser the jet (larger $\kappa$) the sooner (smaller altitudes) recollimation takes place. This can be understood qualitatively by the fact that $\lambda= 1 + q/\kappa$, where $q=|B_\phi/B_z|$ is the magnetic shear at the disk surface (F97). Now, as the mass load $\kappa$ increases, the magnetic lever arm $\lambda$ must decrease (see also Fig.~\ref{fig:LambdaKappa}). This translates into magnetic surfaces that open less, a less efficient magneto-centrifugal acceleration and recollimation shocks that are not only closer to the disk but also with a smaller compression rate $\chi$ due to a smaller FM Mach number $n$. 

However, the physical scales implied are very different. Here, a factor 20 difference in $\kappa$ leads to a decrease of $Z_{shock}$  by a mere factor 3, with a minimum value of 700. In F97, the mass loading parameter $\kappa \sim \xi$ varies from 0.01 to 0.05 only (for a constant disk aspect ratio, see his Fig.~3) but leads to variations in recollimation altitudes that span six decades. Our lowest $Z_{shock}$ obtained for $\kappa=1$ is still much farther away than the minimum height of $\sim 10$ found for $\kappa \sim 0.05$ in F97. 

This discrepancy can of course be attributed to the very different injection properties. Indeed, our numerical simulations assume a supersonic flow while the self-similar calculations compute the disk structure and outflows are found to be only super-SM (and still subsonic) at the disk surface. However, our guess is that the huge difference in the shock position is probably due to the existence of the spine, which breaks down the self-similarity. Indeed, recollimation is due to the dominant hoop-stress and while, in our case, all quantities are leveling off on the axis, strictly self-similar solutions have an axial electric current that grows limitless. For instance, at a cylindrical distance $r=0.1$ from the axis at the spine basis, our $B_z$ remains comparable to the disk field, $B_\phi$ goes to zero and the normalized Bernoulli integral $e$ has decreased by a factor 5 (see Fig.~\ref{fig:Limit conditions}). Self-similar solutions, on the contrary, have fields and a Bernoulli integral increasing respectively by a factor $10^{5/4}=17.8$ and 10. This hints to the fact that the conditions assumed on the axis most certainly affect the overall jet collimation properties. We will come back to this aspect later on by changing the spine properties.

A second interesting aspect is the appearance of a second ensemble of shocks arriving at higher altitudes, namely $Z>3000$ for  $\kappa=0.5$ and $Z>2000$ for $\kappa=1$. This second group of shocks is also composed of two large recollimation shocks separated by smaller ones. The distance (measured at the axis) between the two large shocks is comparable to the width of the first group of shocks. As discussed in Section~3.3.2, this hints to the fact that each group is caused by the global jet recollimation dynamics, that should be periodic in the Z-direction on a scale $H_R$. Looking at the simulation K4 with $\kappa=0.5$, that would give $H_R\sim 1650$, while the width of each group is around $W\sim 300$ (the first group of shocks being located between 1150 and 1450, and the second between 3100 and 3400). 
As long as no dissipation is introduced, such a periodic behavior should continue in a box of infinite size.     

\subsection{Influence of the magnetic field distribution $\alpha$} 

\begin{figure*}[htp]
\centering
\begin{subfigure}{.85\textwidth}
  \includegraphics[height=.3\linewidth]{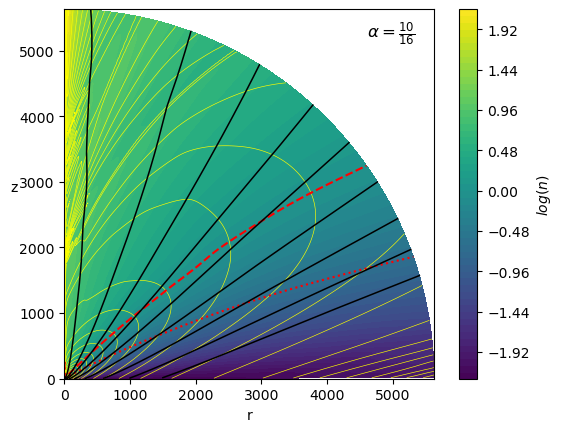}
  \includegraphics[height=.3\linewidth]{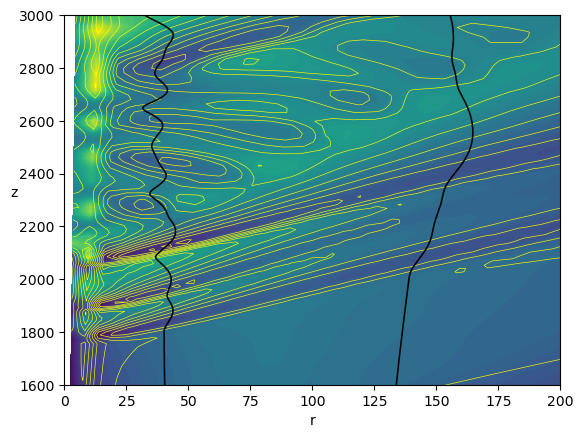}
\end{subfigure}
\begin{subfigure}{.85\textwidth}
  \includegraphics[height=.3\linewidth]{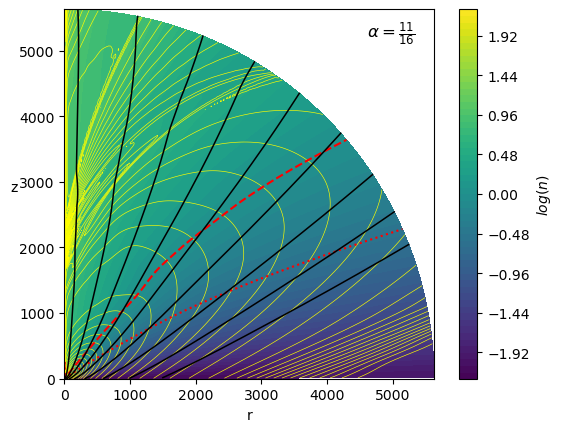}
  \includegraphics[height=.3\linewidth]{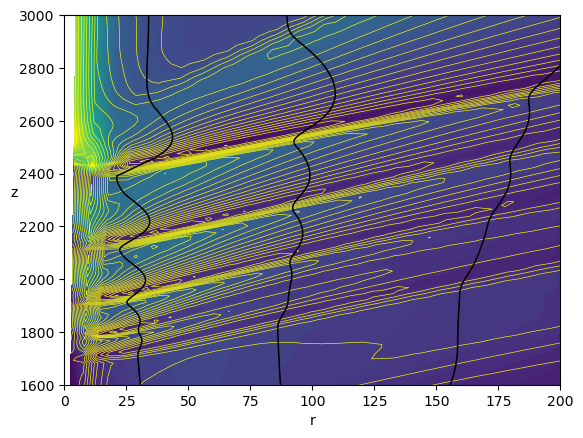}
\end{subfigure}
\begin{subfigure}{.85\textwidth}
  \includegraphics[height=.3\linewidth]{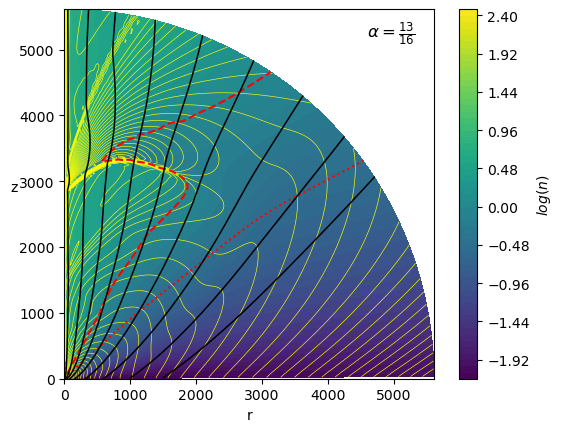}
  \includegraphics[height=.3\linewidth]{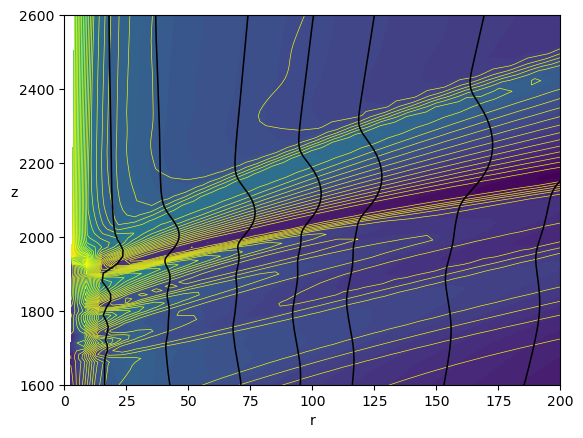}
\end{subfigure}
\begin{subfigure}{.85\textwidth}
  \includegraphics[height=.3\linewidth]{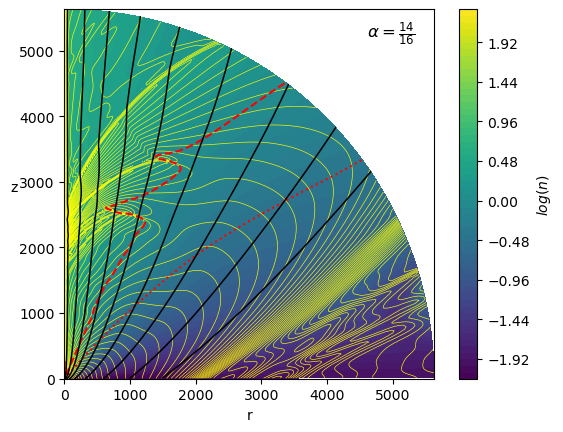}
  \includegraphics[height=.3\linewidth]{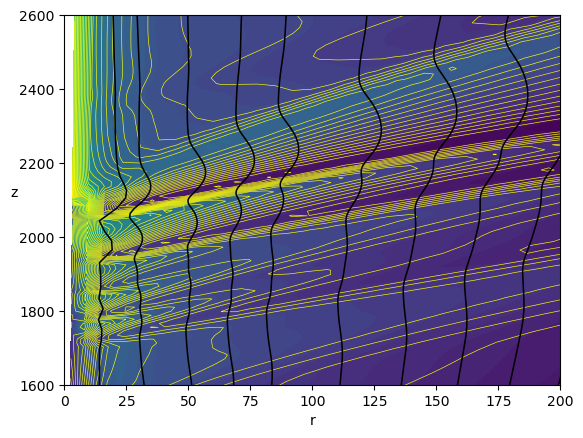}
\end{subfigure}
\begin{subfigure}{.85\textwidth}
  \includegraphics[height=.3\linewidth]{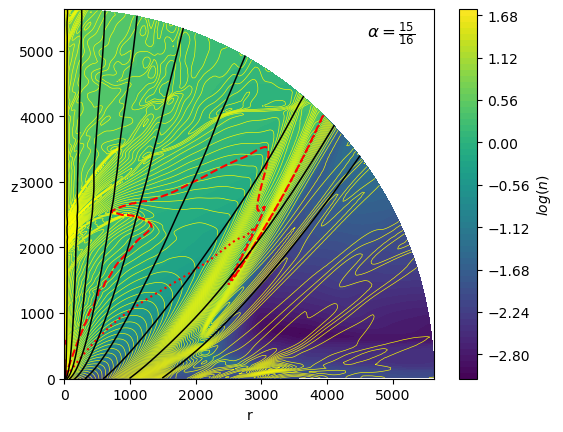}
  \includegraphics[height=.3\linewidth]{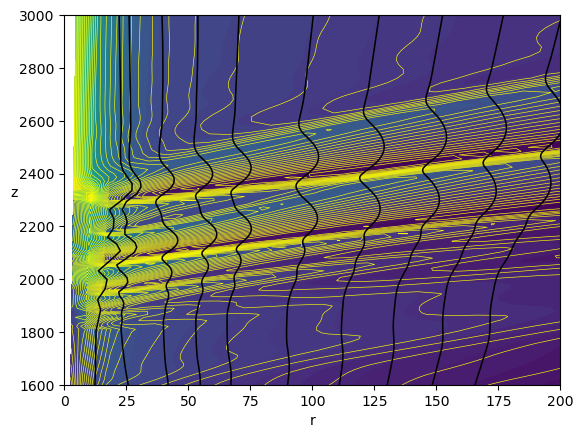}
\end{subfigure}
\caption{Influence of the magnetic field distribution $\alpha$ on the final stage of jets obtained with $\kappa=0.1$. We use the same notations, colors and field lines anchoring radii as in Fig.12}.
\label{fig:SimulationsAlpha}
\end{figure*}

In this section we vary the magnetic flux distribution exponent $\alpha$ in our simulations A1 to A5, keeping all other parameters (see Table~\ref{tab:ParametresSimus}) constant. A strict mathematical self-similarity links the magnetic field distribution $\alpha$ with the disk density in such a way that $\alpha= (12+ 8\xi)/16$, where $\xi$ is the disk ejection efficiency and is related to the disk accretion rate $\dot M_a(r) \propto r^\xi$ \citep{ferreira1995}. As long as material is only outflowing from the disk (namely $\xi >0$) and jet power released only from accretion (namely $\xi <1$, F97), this leads to the unavoidable constraint $12/16 <\alpha < 20/16$. 

Only $\alpha=12/16$ can be compared to the cold jet solutions of F97, assuming an ejection efficiency $\xi <0.1$. Larger values of $\alpha>12/16$ would require a disk ejection efficiency $\xi=0.125$ up to 0.5. These values are only achievable in analytical studies by introducing either an additional heat deposition at the disk surface (magneto-thermal flows, \citealt{casse2000a}) and/or a much smaller magnetic field strength \citep{jacquemin-ide2019}. But the physical processes required to get these solutions are missing in our simulation setup. We will nevertheless vary $\alpha$ in order to allow a comparison with the self-similar jet solutions found by \citet{contopoulos1994}, who played freely with $\alpha$ ranging from 0.5 to 1.02. We did not succeed to numerically obtain steady-state solutions for $\alpha \geq 1$ for reasons that will be discussed below and show only simulations with $\alpha$ ranging from $10/16=0.625$ to $15/16=0.937$. 

All our A1 to A5 simulations reach a steady-state with the same overall behavior as K2, namely the existence of two main recollimation shocks within the computational domain, separated by several smaller standing shocks located at the spine-jet interface. This can be clearly seen in Fig.~\ref{fig:SimulationsAlpha} where a trend with $\alpha$ can however be observed.  Indeed, the radial extension of the shocks decreases with $\alpha$. This can be also seen in Fig.~\ref{fig:KappaShockAltitudeEvolution} which shows that the altitude $Z_{tip}$ of the main recollimation shock decreases with $\alpha$, while its altitude $Z_{shock}$ at the axis barely changes. For simulations A1 ($\alpha= 10/16$) and A2 ($\alpha= 11/16$), the main recollimation shock ends out of the box, thus $Z_{tip}$ cannot be defined. However, the value of $Z_{shock}$ remains similar (see Table 1). This is a geometrical effect due to the fact that, as $\alpha$ increases, the magnetic field configuration goes from a highly inclined ($\alpha=0$ corresponds to a monopole) to a much less inclined magnetic configuration ($\alpha=2$ is a purely vertical field). This can be seen in the shape of the magnetic field lines (black solid lines) in Fig.~\ref{fig:SimulationsAlpha}. This geometrical effect translates into a smaller incidence angle near the axis and therefore to weaker shocks (the incidence becomes normal and $n_\perp$ decreases to unity). 

For $\alpha= 10/16$ and $11/16$, the shocks still exist but the MHD characteristics are much more vertical than in K2. A larger box would probably be necessary to recover the K2 behavior. The opposite trend can be seen for $\alpha= 14/16$, with less vertical MHD characteristics allowing now the second main recollimation shock to merge with the FM surface within the domain.   

Table~\ref{tab:ParametresSimus} also shows that as $\alpha$ increases, the last radius on the disk giving rise to a super-FM flow increases and the colatitudes (measured at the outer boundary) $\theta^{ext}_{FM}$ and $\theta^{ext}_{A}$ of the critical FM and A surfaces decrease. These results are a natural consequence of the magnetic field distribution becoming more vertical as $\alpha$ increases. As the jet mass loss $\dot M_{jet}$ and jet power $P_{jet}$ is computed up to \textbf{$r_{0,FM}$} which increases with $\alpha$, $\dot M_{jet}$ and $P_{jet}$ increase with $\alpha$. Still, the mass loss decreases even when computing up to a fixed radius. Indeed, the density decreases less with an increasing $\alpha$ ($\rho \propto r^{2\alpha-3}$) and the outer disk regions contribute more to the mass flux and jet power. Moreover, as the distribution in density is flatter with an increasing $\alpha$, it is only natural that $\dot M_{spine}/\dot M_{jet}$ and $P_{spine}/P_{jet}$ decrease when $\alpha$ increases.


To summarize, we find that the altitude of the shocks barely changes with $\alpha$, which is in strong contrast with \citet{contopoulos1994}. Indeed,  
their Table~1 shows that as $\alpha$ increases from $0.5$ to a critical value $0.856$, their self-similar jet becomes super-FM and undergoes a recollimation at a distance that increases by several decades (as in F97). As discussed above, we believe that this discrepancy is due to our non strict self-similar scaling (which forbids the unlimited growth of the inner electric current and its Z-pinch in self-similar solutions) and the presence of the spine. They also report that their solutions with $\alpha >0.856$ remain sub-FM, while we clearly achieve super-FM flows up to $\alpha=0.937=15/16$. This is again probably a difference in our jet radial balance, leading to a slightly different jet acceleration efficiency. Finally, their solutions with $\alpha=1, 1.01, 1.02$ remain sub-FM but evolve through a series of radial oscillations at logarithmically equal distances in Z. 


Our simulation A5 is the simulation with the largest value $\alpha=15/16$, close to unity. Although the final integration time $t_{end}$ is quite comparable with the other simulations, its appearance clearly shows that the global configuration is still far from achieving a steady-state like K2. This can be seen in Fig.~\ref{fig:SimulationsAlpha}, in the shape of the critical surfaces but also on the isocontours of the poloidal electric circuit (yellow curves) that are still struggling to find their final state. This is normal and points to a numerical difficulty in computing MHD codes magnetic configurations with $\alpha \geq 1$. 

Indeed, it has been argued above that for self-similar boundary conditions, the jet transverse balance imposes a toroidal magnetic field scaling with the vertical magnetic field. This leads to an electric current at the disk surface behaving as $I=r B_\phi \propto r^{\alpha-1}$. Magnetic configurations with $\alpha<1$ correspond to a poloidal current density leaving the disk surface and closing along or near the spine (where it flows back to the disk), whereas configurations with  $\alpha > 1$ correspond to an inward poloidal current density, with current closure being done only at the outskirts of the outflow (F97). Since $div\,  {\bf J}=0$ in MHD all electric circuits  must be closed. Let us define a radius in the disk $r_I$ such that for $r< r_I$, the electric current flows down into the disk whereas it flows out of it for $r> r_I$. For $\alpha <1$, $r_I$ is always larger but close to $R_d$ implying very short time scales. As discussed in Sect.~\ref{sec:time}, as time is evolved, the outer disk regions provide more current that struggles to reach the innermost disk radius. But since the local time near $r_I$ is small, a global radial balance can be achieved consistently with the electric current closure condition. On the contrary, configurations with $\alpha >1$ have $r_I$ that is constantly increasing in time (as $t^{2/3}$), leading to an electric circuit that freezes in time and therefore to a transverse MHD balance that takes a much longer time to achieve steady-state. We have observed this behavior for all values of $\alpha>1$ and none of these simulations has achieved any steady-state. 

The limiting value $\alpha=1$ (close to our A5 simulation) would correspond to $r_I=R_d$ and absolutely no electric current flowing out of the disk until some outer radius. Current closure could only be done through the spine and the outer jet interface with the ambient medium. But then, no magnetic acceleration would be possible as no electric current could be used along the magnetic surfaces. This would correspond to an exact force-free field configuration fully determined by the chosen boundary conditions. To compute such a situation, boundary conditions with a finite extent of the jet launching region must be designed. However, we doubt that the outcome would be a force-free field unless explicitly enforced. This is beyond the scope of this paper.

\subsection{Influence of a rotating central object}
\label{sec:spine}

In this section, we do not intend to make a full exploration of the physical parameters of the spine but instead wish to probe whether the spine, despite its small spatial extent, has indeed a profound influence on the overall jet dynamics. All simulations  K1-K5 and A1-A5 were done with the same non-rotating central object, in order to minimize its emf and numerically follow the outcome of a jet emitted from an outer self-similar disk (see Sect.~2.4.2). Our choice of parameters gives rise to a spine carrying typically 10\% of the mass flux, being thereby only a small contribution to the overall outflow. Nevertheless, it carries a large fraction of the emitted power, even superior to that of the disk for the simulation A1. The spine plays an important role in introducing extra standing shocks at its interface with the jet, but is probably also determining the altitude where the first large recollimation shock occurs. Indeed, as discussed in the previous sections, the amount of electric current that is flowing along the innermost axial regions (along the spine and the inner jet) is what determines the strength of the Z-pinch acting upon the jet and thereby the altitude $Z_{shock}$.     


In order to probe this idea, we run another simulation with a rotating object (simulation SP in Tab.~\ref{tab:ParametresSimus}). We choose an object rotating at the same angular velocity as the innermost disk radius $R_d$, namely $\Omega_a=\Omega_{Kd}$. This is for instance representative of a star-disk interaction where the disk truncation radius is located at the co-rotation radius. By doing so, the emf due to the central object becomes non-negligible and we expect a stronger poloidal electric current. However, care must be taken as enhancing the hoop-stress may lead also to an overwhelming radial pinch. To prevent this and get somewhat closer to the self-similar conditions, we also increase the value of the Bernoulli integral on the axis and use $e_a=10$ ($e_a=2$ for the other simulations). Note that the Bernoulli invariant from the innermost disk region is $e_d \simeq \lambda_d - 3/2\sim 10$. This translates mostly into a thermal pressure 5 times larger than previously. Thus, our new conditions for the spine provide a rotation and a specific energy that are only comparable to those at the inner jet, not much larger as in a self-similar situation.  


\begin{figure}
\centering
    \includegraphics[trim=0 10 0 17,clip,width=.98\linewidth]{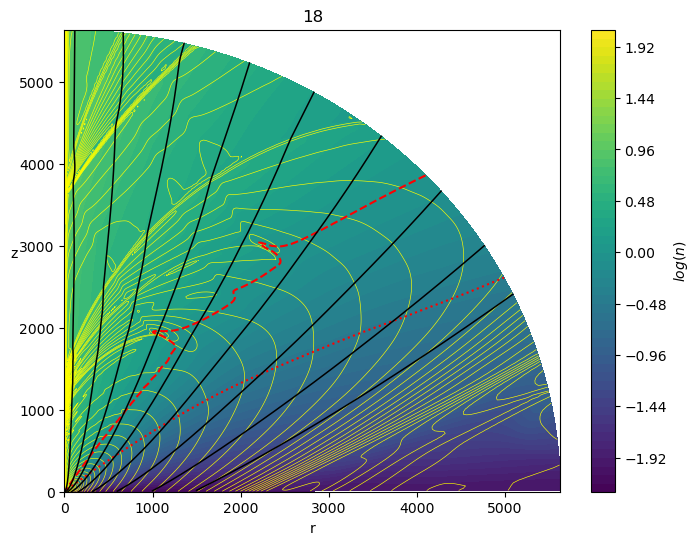}
\caption{Snapshot at $t_{end}$ of our SP simulation with a rotating spine, $\alpha=3/4$ and $\kappa=0.1$. We use the same color coding as in Fig.~\ref{fig:ReferenceSimulation}. The magnetic field lines (black solid lines) are anchored at the same disk radii.} 
\label{fig:Rotation}
\end{figure}

Figure~\ref{fig:Rotation} shows the final outcome of this new simulation. It achieves a global steady-state with the same features as in our reference simulation K2. However, the shocks are as expected localized at lower altitudes, allowing a second set of large recollimation shocks to appear near $Z\sim 3700$. The first large recollimation shock appears at $Z_{shock}=1300$, which is significantly smaller than $Z_{shock}=1850$ obtained for a non rotating central object. This result confirms the role of the central object in shaping, through its spine, the collimation properties of the jets emitted by the surrounding disk. This is very promising and deserves further investigations.

\section{Discussion}
\label{sec:disc}

\subsection{Caveats}

This paper provided some fresh intel on the collimation of jets emitted from self-similar magnetized disks. However, we would like the reader to pay attention to its caveats : 

As we tried to replicate a Blandford \& Payne process, we mostly explored the ejection conditons on the disk and not on the source, with the exception of the rotation (see section 4.3). For instance, we kept a value of sonic mach number $M_S=10$ on the whole ejection zone, thus only creating cold jets. It would be interesting to provide a real exploration of the spine parameters to better understand the role of the central object in the collimation.

The shocks displayed in our simulations are rather weak ($\chi \sim 2$ near the axis, $\chi \sim 1.3$ in the jet). This is probably due to our insentropic scheme, but as the simulations only reach small values of $n$, the compression rates are intrinsically low (see Appendix C). Moreover, as shown in Eq C.4 the impact of those shocks on jet angular velocity in YSOs is probably weak, at least for jets achieving $m^2 >> 1$.

Our adiabatic solutions do not allow energy dissipation, so the shocks should go on, one above the other, each shock producing its own local accelerating circuit. This is showcased by the presence of a second subset of shocks in simulations K4, K5 (see Fig. \ref{fig:SimulationsKappa}) and SP (see Fig. \ref{fig:Rotation}). Thus, this setup does not allow the presence of a "true" asymptotic circuit that would extend up to infinity on the jet axis.

These simulations are highly dependent on the numerical setup. In order to capture the expected shocks, we used an HLLD solver, switching to more diffusive HLL solver and MINMOD linear spatial reconstruction in regions of extremely low density and very high Alfv\'en speed.
For the same reason, we used a higher resolution in $\theta$ around the axis to resolve the shocks. Still, we were only able to reach very long time-scales thanks to a novel method boosting the numerical integration (see Appendix A).

\subsection{Comparison with other numerical works}

In this section, we only compare our findings with previous 2D platform simulations of non-relativistic jets. Indeed, relativistic jets develop an electric force that deeply affects the asymptotic collimation, forbidding a direct comparison with our non-relativistic setup. We also disregard 3D jet simulations as they usually introduce a whole new phenomenology related to jet instabilities that are not present in our work. Making 3D simulations of our jets is planned for future work. 
  
Platform simulations of jets have a lot of degree of freedom and it is therefore very difficult to determine the exact generic results on jet collimation that can be derived from them. Even if the whole injection domain is chosen to be sub-SM, three free distributions must be chosen at the boundary (assumed to be the disk surface), usually $B_z(r),\rho(r)$ and $u_z(r)$. The disk being assumed to be Keplerian, most of the works use field lines rotating at Keplerian speeds and $u_z= V_{inj} V_K$, where $V_{inj}$ is a small dimensionless number. We use the following notations
$ B_z \propto r^{\alpha-2}$ and  $\rho \propto r^{-\alpha_\rho}$ leading to
\begin{equation}
\kappa = \frac{V_{inj}}{\mu^2}= \kappa_d r^{\alpha_\kappa} \ ,   \, \,\   \mu= \frac{V_{Az}}{V_K} =\mu_d r^{\alpha_\mu}
\end{equation} 
with $\alpha_\kappa= -2\alpha_\mu = -2\alpha +\alpha_\rho + 3$ and the cylindrical radius is normalized to the inner radius $R_d$. For a given magnetic field distribution, the way the mass is injected in the outflow (or how the magnetic energy must vary within the disk) is an important quantity allowing to compare the various jet models. Our injection conditions (Eq.~\ref{eq:InjectionCondition}) have $\alpha_\rho=2\alpha-3$ thus $\alpha_\mu=\alpha_\kappa= 0$ in agreement with self-similar studies (BP82, F97). For our explored range in $\alpha <1$, our jets are always dominated by the mass flux emitted from the innermost disk regions.  As discussed previously, the spatial and temporal scales are also very important in order to obtain recollimation shocks. We recall that our spherical domain goes up to $R_{ext}=5650R_d$ covered with $266\times 1408$ zones and lasted $T_f \ga 10^5\, T_d$. 

The box size and time scales achieved in the pioneering works of \citet{ouyed1997,ouyed1997a,ouyed1999} were of course quite small, with a cylindrical domain $(z,r)=(80,20)$ in $R_d$ units and a resolution of $(500,200)$ cells, the simulations lasting up to $T_f\sim 500$. They studied mostly $\alpha=1$ and $\alpha=2$ magnetic configurations, assuming $\alpha_\rho=3/2, \mu_d=0.01$ and with no injected spine. Their jets have therefore a steeply decreasing $\kappa$ (or increasing $\mu$) with $\alpha_\kappa=3/2-2\alpha$, providing situations very different from ours. They argued that the nature of the outflow (steady or not) is mostly determined by the mass load $\kappa$, with unsteady jets containing shocks and associated knots arising at small values of $\kappa_d\sim 10^{-2}$. While these shocks are indeed due to jet material being focused toward the axis, no steady-state situation is reported in their case. Our own simulations show that their time scales were still too short to warrant a transverse jet balance, especially for $\alpha \geq 1$. Moreover, it remained unclear whether these knots were indeed a consequence of a small mass load $\kappa_d$ or due to the boundary conditions used at the jet basis, which were too numerous and thereby over-determined the outflow dynamics (see discussions in e.g \citealt{bogovalov1997, krasnopolsky1999, ramsey2019}).

\citet{ustyugova1999} showed that if the simulation region is elongated in the $z$-direction, then Mach cones may be partially directed inside the domain, leading to an artificial influence (usually collimation) on the flow. Using a domain $(z,r)=(200,170)$ with  \textbf{$100 \times 100$} cells, they showed that this effect can be reduced with a square or spherical grid.

\citet{pudritz2006} extended their work by exploring a larger range in $\alpha=1, 3/4, 1/2, 1/4$, using $\kappa =5 r^{3/2-2\alpha}, \mu_d=0.01$ and no spine. They argued that the collimation of a jet depends on its radial current distribution, which in turn is prescribed by the mass load. Simulations with $\alpha=1, 3/4$ would collimate to cylinders due to a decreasing $\kappa$ leading to a large $B_\phi$, whereas simulations with $\alpha=1/2, 1/4$ with an increasing $\kappa$ would produce a smaller $B_\phi$ and jets closer to wide-range outflows with parabolic collimation. However, our simulations show that the physical scales needed to observe the correct asymptotic state are much larger than those achieved in these early simulations. Moreover, it is indeed correct that self-collimation depends on $B_\phi$ which, in a magnetic jet that carries away the disk angular momentum namely $-rB_\phi/\eta \propto \Omega r^2$, varies as $B_\phi \propto \kappa(r) B_z$ at the disk surface. In their case, this expression leads to $B_\phi \propto r^{-1/2-\alpha}$, which is indeed more steeply decreasing with $\alpha$. Our guess is however that the collimation observed within their box is mostly a consequence of the potential magnetic field configuration used as the initial condition, as also illustrated in Fig.~\ref{fig:EvolutionFieldLines}. Therefore, the smaller $\alpha$ and the wider (less collimated at a fixed distance) is the jet.     

The influence of the magnetic field profile $\alpha$ on the asymptotic jet collimation has also been investigated by \citet{fendt2006}, who performed 40 simulations in a larger cylindrical grid $(z,r)=(300,150)$ with $256 \times 256$ cells, the simulations lasting up to $T_f\sim 300$ to $5\, 10^3$ (for those achieving a steady-state over at least 50\% of the grid). He explored a wide range in $\alpha_\rho$ from 0.3 to 2 and in $\alpha$ from 0.5 to 1.8, using the same boundary conditions as \citet{ouyed1997a}, with $V_{inj}=10^{-3}, \kappa_d=5$, no spine and $\mu_d$ varying between 0.1 and 2.67. He confirmed that the degree of collimation is decreasing for a decreasing $\alpha$ regardless of $\alpha_\rho$, in agreement with our suspicion that the overall MHD collimation trivially follows the potential field configuration (see also Sect.~3.2). For $\alpha > 1.6$, no steady-state jet is actually found, with a wavy radial pattern evolving along the outflow. This is both consistent with our finding that for $\alpha\geq 1$ the time scales for reaching stationarity become overwhelmingly long and the existence of radially oscillating, sub-FM analytical solutions for $\alpha\geq 1$\citep{contopoulos1994}. \citet{fendt2006} also reports a degree of jet collimation increasing with the jet magnetization exponent, namely with $\mu_\sigma= -\alpha_\kappa -3/2$ (see his Eq.~10). Now, of the 40 simulations only 6 have $\alpha_\kappa=\alpha_\mu=0$ and 8 have $\alpha_\kappa>0$, so that most simulations describe a mass loading decreasing with the radius. None of the simulations show standing recollimation shocks, even in the BP82 case obtained with $\mu_d= 0.177$ (we use $\mu_d=1$). Putting aside this difference, our Fig.~\ref{fig:EvolutionShockPosition} shows that around their final time $T_f=5000$, we observed shocks only around $Z\sim 1900$, far beyond their computational domain. 

\citet{krasnopolsky1999} used a cylindrical grid $(z,r)=(80,40)$ with $256 \times 128$ cells, the simulations lasting up to $T_f\sim 170$, introducing a ballistic axial flow below $R_d$ (the spine), injected close to the escape speed and surrounded by a disk wind. They used the correct number of boundary conditions and, by playing with the size of the box, they showed the drastic importance on the overall flow collimation of the amount of magnetic flux becoming super-A within the box. They studied mostly $\alpha=1/2$ and 3/4 with $\mu_d=4$ and rather flat density distributions leading to $\alpha_\kappa>0$, from 2 to 3/2. They do not report any time-dependent behavior seen in previous studies, which they attribute to both the existence of their sub-FM inner spine (where magneto-centrifugal acceleration is inefficient) and the correct treatment of boundary conditions. 
This work was extended by \citet{krasnopolsky2003} on a much larger box $(z,r)=(10^3,10^3)$ with $190 \times 210$ zones, the simulations lasting an unspecified time $T_f$. They only studied the case $\alpha=1/2$, with ejection from a finite zone $r_o=R_d$ and $r_o=10R_d$, yielding $\alpha_\rho=1$ ($\alpha_\kappa=1$) or $\alpha_\rho=3$ ($\alpha_\kappa=-1$). They found that the collimation degree of this finite jet is improved for a steeper density profile, namely with a decreasing mass load with the radius, as discussed above. They report neither recollimation toward the jet axis nor radial oscillations and attributed this behavior to their non self-similar scaling. Our own results show instead that recollimation should be seen farther out (beyond their box) and that radial oscillations are expected only for $\alpha>1$.  

Using the same grid and numerical setup as \citet{krasnopolsky2003}, \citet{anderson2005} studied the effect of $\kappa_d$ on the collimation of a cold BP82 jet model with $\alpha=3/4$ and $\alpha_\rho=3/2$ (thus $\alpha_\kappa=0$). They varied $\kappa_d$ from $6.3\, 10^{-3}$ to $19$ assuming that ejection takes place only from $r_o=R_d$ and $r_o=10\, R_d$ (enforcing however $B_z$ to zero at the edge of the launching region), while we assumed ejection from the whole disk and varied $\kappa_d$ only from $5\, 10^{-2}$ to 1. Despite the truncation due to limited ejection range and the (almost) purely radial magnetic field at the edge of the launching region, they recover the same results as in steady-state jet theory (FP97): jets get more and more open as $\kappa_d$ decreases (see discussion in Sect.~4.1). They do not report any recollimation shock (although wiggles can be seen in their Fig.~4) but again, our shocks fall below $Z=1000$ (within their box) only for $\kappa_d\sim 1$ (see Fig.~\ref{fig:KappaShockAltitudeEvolution}). We conclude that their box was too small to observe any standing recollimation shock. \citet{anderson2005} report the inability to reach steady state (the time scale $T_f$ is unspecified) for $\kappa_d$ larger than unity, when field lines start to oscillate and produce ripples that propagate outward. This behavior is consistent with analytical studies and is related to the capability to produce super-A flows when they are heavily loaded (or have a weak magnetic field). Indeed, magnetically-driven cold flows are possible only up to $\kappa \sim 1$, leading to a magnetic lever arm $\lambda \sim 2$. For larger mass loads (and smaller $\lambda$) gravity plays an important role, with the Alfv\'en surface getting closer to the disk, requiring the field lines to be bent by much more than the fiducial $30^\circ$ at the disk surface\footnote{This is the reason why we could not reach steady-state solutions with $\kappa >1$ with our setup.} (see Fig.~4 and discussion around the Grad-Shafranov equation in \citealt{jacquemin-ide2019}).

The largest axisymmetric simulations have been provided by \citet{ramsey2011,ramsey2019}, using 9 levels of AMR in a  cylindrical grid $(z,r)=(8\, 10^4,5\, 10^3)R_d$ with simulations lasting up to $T_f\sim 6\, 10^4$. They computed the propagation and evolution of 8 jets up to observable scales, defined with varying mass loads $\kappa_d$ from $5\, 10^{-2}$ to 32 and $\alpha_\rho=3/2, \alpha=1$ (thus a decreasing mass load with $\alpha_\kappa=-1/2$). Mass is injected with $V_{inj}=10^{-3}$ and there is no injected spine as in \citet{ouyed1997}, although a spine naturally emerges. In all simulations, they observe that regions beyond $r_o\sim 10\, R_d$ fail to displace the hot atmosphere and that the outflow is stifled, despite the decrease of $\kappa$. This is actually consistent with our previous discussion for simulations with $\alpha\geq 1$, which take a much longer time scale to reach steady-state. Nevertheless, as the inner parts of the outflow evolve on much shorter time scales, some quasi-stationary situation can settle (see their Sect.~5.3). With no surprise, this is the case for small mass loads, while knots appear for $\kappa_d=0.5$ (simulation E) and are recurrent (quasi-periodic) for $\kappa_d=2$. These knots are not to be compared with our standing recollimation shocks, as none of the MHD invariants are constant along field lines passing through them. They are made of plasmoids launched from \textbf{$R_d \la r_o \la 2 R_d$}, where gas is both dense and hot. The knot formation mechanism is here directly related to the jet launching process from this innermost disk region. Indeed, in this region the field line bending is insufficient to drive the massive injected material, until a sufficiently strong toroidal field builds up and lifts the matter, in agreement with steady-state theory of massive outflows (F97, \citealt{jacquemin-ide2019}). The regularity of knot spacing is indicative of a simple oscillator related to the necessary build-up of a strong toroidal field. These plasmoids are magnetically confined by the surrounding poloidal magnetic field, follow the path of the jet and eventually merge together. For larger mass loads ($\kappa_d=8$ and 32, simulations G and H), the outflows are fully unsteady while keeping their global structure (probably because of their 2D nature, since destroying instabilities such as kink or Kelvin-Helmholtz require 3D, as argued by the authors).  

To our knowledge, no previous jet simulation has shown the existence of standing recollimation shocks, either because the computational domain was too small and/or the simulation time scales were too short. These limitations are even worse of course for simulations that do take into account the disk physics, as they must also struggle to follow the disk and the mass loading process. 

The first of these simulations computed an accretion-ejection configuration with $\alpha=3/4$ and $\alpha_\rho=3/2$ (the BP82 case) within a cylindrical grid $(z, r)=(80, 40)$ on a time $T_f=251$ only \citep{casse2002,casse2004}. On these time scales, the mass loading process is computed, leading to the inside-out establishment of self-similar conditions with $\alpha_\kappa=0$. Further simulations, done also with the same initial configuration but exploring various disk parameters, were computed on slightly extended scales, a grid $(z, r)=(120, 40)$ on a time $T_f=400$  \citep{zanni2007,tzeferacos2009, tzeferacos2013} and a grid $(z, r)=(180, 50)$ on a time $T_f=5.6\, 10^3$ \citep{sheikhnezami2012}. Since most of these works were focused on the disk physics and less on the jet dynamics, not much is shown about the latter. The simulations of \citet{stepanovs2016} were done on a spherical grid up to $R_{ext}=1500$ with $(N_R \times N_\theta)=(600\times 128)$ zones and up to $T_f=10^4$, for the same BP82 initial configuration. Such scales would be relevant for the appearance of recollimation shocks but they only show close-up views below $R=30$ and focus instead on the accrection-ejection correlations. However, the long time scales allow to see a radial redistribution of both the vertical magnetic field and the disk density (since both evolve on accretion time scales \citealt{jacquemin-ide2019}), thereby modifying the initial strict self-similar conditions. 

The time evolution of the disk magnetic field distribution has been reported previously \citep{murphy2009}. These simulations were done in a cylindrical grid $(z, r)=(120, 40)$ up to a time $T_f\simeq 6\, 10^3$, and using $\alpha=1/4$ with $\alpha_\rho=3/2$. Such an initial magnetic field distribution leads to a magnetic energy density on the disk midplane decreasing very rapidly ($\propto r^{-1}$), so that a super-FM ejection (with proper MHD invariants) takes place only up to some radius $r_o\sim 5$ \citep{murphy2010}. Their paper was focused on this ejection from a limited zone within the disk and not much was said about the jets. However, we report that on the long time scale of the simulation, the magnetic field is seen to slowly evolve within the disk leading to some readjustments also in the jet. How such a modification affects the jet transverse balance and possible standing recollimation shocks is an open issue that deserves further investigations.

%
Note that standing recollimation shocks have been already discussed in steady-state 2D jet simulations built upon analytical self-similar solutions. In these works, a cylindrical box is used which starts at a $z_o$ well above the disk (say $z$ from $z_o=10$ to 210 and $r$ from 0 to 100 in units of $R_d$). This allows to fill in the whole domain with either only a self-similar BP82 jet model \citep{gracia2006, stute2008}, or a combination of an axial (meridionally self-similar) stellar wind surrounded by a BP82 jet model \citep{matsakos2008, matsakos2009}. The numerical procedure, which evolves in time the MHD equations for a set of boundary conditions, allows to rapidly obtain a stationary solution on time scales $T_f\sim 40$ to $10^3$. A weak recollimation shock is always found between the axial flow and the BP82 jet, which fulfills most properties discussed in our paper. However, in strong contrast with our own work, the existence of this shock is unavoidable in these works and directly imposed by the boundary conditions. Indeed, the outflow is already super-FM at the injection altitude $z_o$ for all radii below $r_o \sim 6$ (see for instance Fig.~1 in \citealt{matsakos2008}), while field lines are already being focused toward the axis.   

\subsection{Astrophysical consequences}

In this paper, we showcase one mechanism enabling the creation of a recollimating jet and its subsequent shocks. There are other models explaining the creation of such shocks. They could be triggered for instance by a sudden mismatch between the jet and the ambient medium pressure. Studying FRII jets such as those from the radio galaxy Cygnus A, \cite{komissarov1998} proposed that the jet confinement and its consequential shocks are caused by the thermal pressure of an external cocoon. For the case of FRI jets, in \cite{perucho2007} the jet expands until it becomes under-pressured with respect to the ambient medium, then recollimates and generates shocks, unless a turbulent mixing layer at its interface with the ambient medium forbids its formation \citep{perucho2020}. In any case, such shocks happen much farther away than in our case and depend critically on the ambient pressure distribution.

On the contrary, the jets in our simulations are intrinsically collimated by the self-induced hoop stress (see Fig. \ref{fig:RadialForces}). As shown in FP97 for self-similar cold models and proved here in full 2D time-dependent simulations, this force will lead the cold jets toward the axis, leading to the formation of standing recollimation shocks. Such a mechanism should therefore apply regardless of the external medium and around various astrophysical objects.


Extragalactic jets imaged by VLBI display knots of enhanced emission that could be associated with shocks (as they play an important role for the production of non-thermal emission). While most of these features are moving, some of them appear stationary (\citealt{lister2009,lister2013,doi2018} and \citealt{boccardi2017} for a review). The closely studied M87 jet is a particularly interesting case. It contains several moving and stationary bright features near the HST-1 complex (\citealt{asada2012, walker2018, park2019}), whose origin may be due to pressure imbalance when the jet reaches the Bondi radius. This distance is actually larger than the scales reached by our simulations. However, these are  newtonian and it is unclear if relativistic effects (in particular the decollimating force due to the electric field) would not push farther out the recollimation scale. In any case, we note that our non-relativistic simulations provide shocks that are located on the same scale as the closest features in the M87 jet (see figure 2 of \citealt{asada2012}).


Protostellar jets also present some interesting features along the flow usually interpreted as being bow shocks, like in HH212 (\citep{lee2017}) or HH30 (\citep{louvet2018}). Their origin remains highly debated, either instabilities triggered during jet propagation or variability induced by a time-dependent jet production mechanism (as advocated for instance in HH212 by the remarkable jet-counter jet symmetry, see \citealt{tabone2018}). We suspect however that whenever a jet undergoes an MHD recollimation shock, that refracts the jet away from the axis, then more shocks are to be expected downstream (and probably affected by the external pressure distribution). MHD recollimation may therefore provide an intrinsic means to trigger jet variability on observable scales. Stationary emission features are sometimes indeed detected, as in HH154 \citep{bonito2011}. These features are located from a few tens to a few hundreds of $au$ from the source, a distance comparable to the altitude of the first standing recollimation MHD shock. This should be worth further investigation.

\section{Conclusion}

We present axisymmetric simulations of non-relativistic MHD jets launched from a Keplerian platform. These are the first to show the formation of standing recollimation shocks, at large distances from the source. These recollimation shocks are intrinsic to the MHD collimation process and have been proposed as a natural outcome of self-similar jet launching conditions (F97, \citealt{polko2010}). Because they were never seen in previous MHD simulations of jets, the suspicion grew that recollimation would be a bias due to the self-similar ansatz. It turns out that the physical scales required to capture these shocks are much larger than those used in previous works. Using unprecedented large space and temporal scales allowed us to firmly demonstrate the existence of such internal standing shocks and, thereby, to bridge the gap between analytical and numerical approaches.   

We analyse the conditions of formation of these recollimation shocks and show that they qualitatively follow the behavior demonstrated in analytical studies, namely that they get closer to the source as the mass load increases. We also confirm that the magnetic field distribution in the disk ($B_z \propto r^{\alpha-2}$) is the key quantity shaping the asymptotic jet collimation. For our self-similar ejection setup, this MHD collimation closely follows the trend satisfied by the potential field: the larger the $\alpha$ the stronger the collimation. However, no steady-state solution has been obtained for $\alpha \geq 1$, because of the difficulty in establishing a stationary self-consistent poloidal electric circuit along the outer jet regions. Since the magnetic field distribution is very likely to evolve on the accretion time scale, we expect jet signatures to vary as well (see e.g. discussion in \citealp{barnier2022}).     


Despite their qualitative agreement with analytical studies, our results show an undeniable impact of the central axial flow on the jet asymptotics. This inner spine is not related to the Keplerian disk but instead to the central object and its interaction with the surrounding disk. Indeed, the spine carries a poloidal electric current responsible for the innermost jet collimation. But it may also introduce extra localized spine-jet interactions, leading potentially to disruptive instabilities (like kink and/or Kelvin-Helmholtz) or, on the contrary, to global jet stabilization in 3D. Going to 3D is therefore a necessity to assess its role and the possible persistence of recollimation shocks. Anyway, our results confirm the role of the central object in shaping, through its spine, the collimation properties of the jets emitted by the surrounding disk. This is a very interesting topic that will deserve further investigations.


These internal recollimation shocks introduce several interesting features: (i) an enhanced emission likely seen as stationary knots in astrophysical jets; (ii) a sudden decrease in the rotation rate of the ejected material and (iii) a possible electric decoupling between the pre-shock and the post-shock regions. This is especially of great interest since these shocks occur at observable distances, typically $\sim 150-200$ au in the case of a YSO. However, our setup also assumes ejection up to several hundreds of au, which is clearly inconsistent with derived jet kinematics (see e.g. \citealt{ferreira2006,tabone2020} and references therein). Simulations with ejection from only a finite zone within the disk (the JED) must therefore be done in order to verify whether MHD recollimation shocks are indeed maintained. This is a work in progress.
  

\begin{acknowledgements} 
We  thank  the  referee  for  providing  thoughtful  comments on the manuscript. The authors acknowledge financial support from the CNES French space agency and PNHE program of French CNRS. All the computations presented in this paper were performed using the GRICAD infrastructure (https://gricad.univ-grenoble-alpes.fr), which is supported by Grenoble research communities.
\end{acknowledgements}

\bibliographystyle{aa}
\bibliography{references5.bib}

\begin{appendix}

\section{Long time scales evolution}

\indent \par In numerical simulations, the time increment is fixed by the Courant–Friedrichs–Lewy condition $\Delta t < \Delta x/C_{max}$, where $C_{max}$ is the maximal wave speed in the cell (in our case $u+v_{FM}$) and $\Delta x$ is the cell size. 
In a standard simulation, the time increment of all cells is chosen by taking the absolute minimum of all the time increments in the full computational domain. In our simulations this time step is set by the smallest cells around the inner spherical boundary at $R_d$, which also happen to have the strongest field and the highest Alfv\'en speed.
 However, these cells near the source are also the ones that converge the fastest to a stationary solution. Thus the cells that we consider are converging to a steady state (meaning the relative variation of the density in one integration step is smaller than an arbitrarily small parameter) are not used to determine the time increment of the cells that are still evolving in time. The time increment used to evolve the evolving cells is computed by taking a minimum over only the cells that have not converged yet. The cells that have converged to a steady state are still integrated in time using their own local time increment so as to ensure the stability of the integration and to be able to capture any perturbation that could possibly alter their steady condition. Since the cells that converge the fastest are those characterized by the shortest time increment, the time increments used to evolve the cells that are still evolving in time and have not converged yet to a steady state becomes larger and larger.
 It is important to point out that the stationary solutions obtained with this time boost are also a solution of the standard non-accelerated algorithm.
 
\begin{figure}
\centering
\includegraphics[width=.98\linewidth]{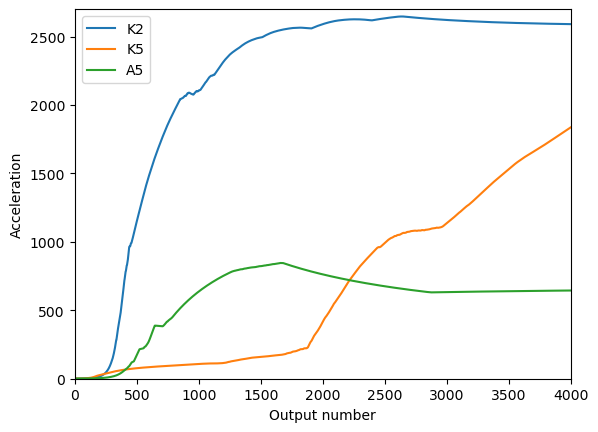}
\caption{Evolution of the acceleration for the simulations K2, K5 and A5.}
\label{fig:TimeEvolution}
\end{figure}

 In Figure \ref{fig:TimeEvolution} is represented the gain in computing time obtained thanks to the time boost. The acceleration factor, defined as the ratio between the physical time reached using the time boost and the physical time that would have been achieved using the standard CFL condition, is plotted versus the progressive numbering of the outputted files.
 Without the acceleration due to our handling of the CFL condition, the time interval between two outputs would have been constant, and the physical time of the solution would have been proportional to the output number.  Any increase of the acceleration factor means that another batch of cells has converged. That means the time  increment of the cells that are still evolving in time becomes larger, thus increasing the reached time scales.
 This increase is clearly visible after the 300th output.
 Using the output number as a proxy for the computational cost of a simulation, this figure clearly shows that at the end of the integration the time boost enables us to reach time scales at least two or three orders of magnitude larger than using a standard CFL condition, without increasing the computational cost of the simulation. Analogously, without employing the time boost we would have required two or three orders of magnitude more CPU hours to reach the same time scales.
 Our approach enabled us to produce simulations that would have consumed way more computing time otherwise. The reference simulation K2 consumed 725 CPU hours, but without the time boost it would have required almost two million CPU hours. It enabled us to work on simply 64 processors kindly provided by GRICAD (Grenoble Alpes Recherche - Infrastructure de Calcul Intensif et de Données).
 
 For the evolution of the acceleration factor with the mass load, we can see in Table \ref{tab:ParametresSimus} that the simulations with $\kappa$ the closest to that of \cite{blandford1982} converge the fastest, reaching larger timescales at the end of the computation. Nevertheless, all seem to reach comparable convergence speeds: in Figure \ref{fig:TimeEvolution} we see that the simulation K5 reaches an acceleration factor similar to that of K2 at the final output. For the evolution with the magnetic field, we can clearly see that the higher the $\alpha$ and thus the flatter the profile of the vertical magnetic field, the slower the simulation converges. For higher values of $\alpha$, the jet is initially more collimated as $B_r/B_z$ is higher on the disk. Thus the field lines further on the disk have a higher impact, retarding the global convergence. That is why the simulation A5 with $\alpha=15/16$ has not converged yet. As an instability develops, some cells that were previously stable become unstable, hence a decrease in acceleration.

\section{Boundary conditions}

\par
\begin{figure}
\centering
  \includegraphics[width=.95\linewidth]{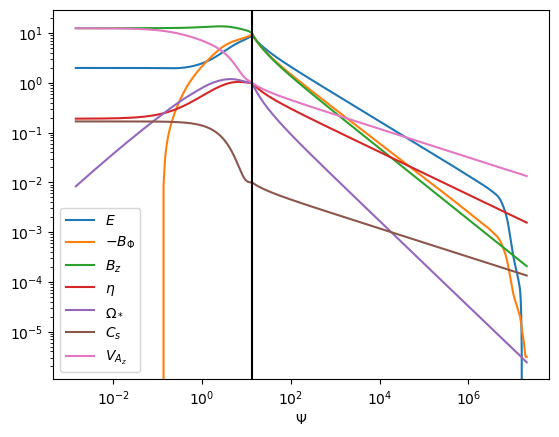}
\caption{Conditions on the lowermost cells of the reference simulation K2. Here all parameters are traced along the first cell above the lower boundary : $R=1 \text{ and } \theta \in [0;\pi/2]$  on the source and then $\theta = \pi /2 \text{ and } R \in [0;R_{ext}]$ on the disk. Are represented the Bernouilli invariant $E$, the toroidal and vertical magnetic field $B_{\Phi}$ and $B_z$, the mass to magnetic flux ratio $\eta$, the rotation speed of magnetic surfaces $\Omega_*$, the speed of sound $C_s$ and the vertical Alfv\'{e}nic speed $V_{A_z}$, over the magnetic flux $\Psi$. The black vertical line corresponds to the flux anchored at ($R=1$,$\theta=\pi/2$), at the source/disk interface.}
\label{fig:Limit conditions}
\end{figure}

 The plots represented in Figure \ref{fig:Limit conditions} illustrate the injection boundary conditions we have chosen, for the reference simulation K2. The Bernouillli invariant $E$, the poloidal magnetic field $B_{\Phi}$, the vertical magnetic field $B_z$, the mass to magnetic flux ratio $\eta = \mu_{0} \rho v_p / B_p$, the rotation speed of the magnetic surfaces $\Omega_{\star} = \Omega - \eta B_\Phi /(\mu_0 \rho r)$, the speed of sound $C_s=\sqrt{\gamma P / \rho}$ and vertical  Alfv\'{e}nic speed $V_{A_z} = B_z / \sqrt{\mu_0 \rho}$ are plotted on the first cell over the injection boundary. The toroidal magnetic field goes to zero on the axis for symmetry reasons, and $\textbar B_{\Phi}/B_z \textbar \gtrsim 1$ on the disk : The JED magnetic field is weakly toroidal. The launching conditions are very cold as $V_{A_z}/C_s \sim 10^2$ on both the source and the disk.

\par The reader may observe the power law dependency with the magnetic flux on the whole disk ($\Psi>10$, after the black vertical line) for all parameters but the electric current. This is directly induced by the self-similar ansatz. However, the torroidal magnetic field $B_{\Phi}$ breaks the power law dependency and shows a swift decrease for $\Psi>2.10^6$. Of all eight variables, only $B_R$ and $B_\Phi$ are free at the injection boundary (see Equation (\ref{eq:InjectionCondition})), and need to cross a characteristic surface to be fixed. For the toroidal current it is the Alfv\'{e}nic surface. As all magnetic surfaces over $\Psi \gtrsim 10^9$ never cross the Alfv\'{e}nic surface, the current can never be fixed. Thus, the simulation cannot ever be stationary in this region.

\par As the disk is a self-similar Jet-Emitting Disk, all dimensionless parameters are assumed to be independent from the radius (\cite{blandford1982}). All these parameters are regrouped in section 2.4.

\par As explained in section 2.3.2.2, we defined a spline function $f(\theta)$ equal to zero on axis ($\theta = 0$) and one at the inner disk radius $(\theta = \pi/2)$ to smoothly connect the axis values with the inner disk ones : \textbf{$f(\theta) \equiv (3 \sin^2\theta - 2 \sin^3\theta)^{3/2}$}.


\par The injection speed ($V_R$ on the source, $-V_\theta$ on the disk) is fixed at $\kappa \mu^2=0.1$ even when varying $\kappa$. However, even though the injection Mach number $M_s=u_p/C_s$ is assumed to be the same all along the boundary, in order to account for the varying inclination of the magnetic surfaces, its value is modified with the parameter $\alpha$ from simulation to simulation. Its variation with $\alpha$ is shown in Table 1.

\section{Rankine-Hugoniot jump conditions}

 
In this section, we write the Rankine-Hugoniot jump conditions valid for standing, adiabatic, recollimation shocks. Contrary to \citet{ouyed1993}, we take into account the toroidal magnetic field since the shocks arise when that component is dominant. The local jump $[A]=A_2-A_1$ between a pre-shock quantity $A_1$ and its post-shock value $A_2$ are expressed in the rest frame as 
\begin{eqnarray}
    [\rho u_\perp ]  = 0 &&  \nonumber \\
    [\rho u_\perp (\frac{u^2}{2} + H) + \frac{B^2}{\mu_o}u_\perp - \frac{\vec u \cdot \vec B}{\mu_o} B_\perp ] = 0 && \nonumber \\
   [P + \rho u_\perp^2  + \frac{B^2_\parallel - B^2_\perp}{2 \mu_o} ] =0 &&  \nonumber \\
   [ \rho u_\perp \vec u_\parallel - \frac{B_\perp}{\mu_o} \vec B_\parallel  ]  =0 && \\
    [ B_\perp ] = 0  &&  \nonumber \\
  [ B_\perp \vec u_\parallel - u_\perp \vec B_\parallel ] = 0  &&  \nonumber 
\end{eqnarray}
where $H=C_s^2/(\Gamma-1)$ is the enthalpy and $u_\perp, B_\perp$ (resp. ${\bf u}_\parallel, {\bf B}_\parallel$ are the normal (resp. tangential) components to the shock surface. The shock is axisymmetric so the tangential component of the magnetic field ${\bf B}_\parallel = B_t \vec e_t + B_\phi \vec e_\phi$, whereas the poloidal component is ${\bf B}_p = B_t \vec e_t + B_\perp \vec e_\perp$, with the unit vectors $(\vec e_\perp, \vec e_t, \vec e_\phi)$ defining a local orthonormal coordinate system. Since these jump conditions express the conservation of mass, angular momentum and energy in ideal MHD, the five MHD invariants along a given magnetic surface ($\eta, \Omega_*, L, E, S$) are thus obviously also conserved (see Fig.\ref{fig:Invariants}). 

In the case of a shock, the mass flux through the surface is non zero, which requires $B_{\perp 2}= B_{\perp 1}\neq 0$ and leads to 
\begin{equation} 
\vec B_{\parallel 2} = \frac{m^2 - 1}{m^2/\chi  - 1} \vec B_{\parallel 1} 
\end{equation}
where $m= u_\perp/V_{Ap,\perp}= u_p/V_{Ap}$ is the Alfv\'enic Mach number of the incoming (pre-shock) flow and $\chi= \rho_2/\rho_1= u_{\perp 1}/u_{\perp 2}$ is the shock compression rate. This equation shows that there are three non-trivial discontinuities with $\chi \geq 1$: (1) an oblique shock with $m^2>\chi >1$, (2) a normal shock with $m^2=\chi>1$ (requiring $\vec B_{\parallel 1}=0$) and (3) an Alfv\'en shear discontinuity with $m^2=\chi=1$ (allowing an arbitrary jump between the two tangential field components). The oblique shock is the only case studied here. 

After some algebra all post-shock quantities can be expressed as function of the pre-shock ones, in particular
\begin{eqnarray}
&& \frac{B_{\phi 2}}{B_{\phi 1} }  =  \frac{B_{t 2}}{B_{t 1} }  = \chi \frac{m^2 - 1}{m^2 - \chi}   \nonumber \\
&& \frac{u_{\phi 2}}{u_{\phi 1} } = \frac{m^2-1}{m^2-\chi} \frac{m^2 r_A^2 - \chi r^2} {m^2 r_A^2 - r^2}  \label{eq:uphi} \\
&&\frac{P_2}{P_1} = 1 + \Gamma m^2_s (\chi -1) \left (  \frac{1}{\chi } + \frac{b^2}{2} \frac{ 2\chi  - m^2(1+\chi )   }{ (\chi -m^2)^2} \right ) \nonumber \\
&& \frac{T_2}{T_1} =  \frac{1}{\chi } \frac{P_2}{P_1}  \nonumber 
\end{eqnarray}
where the sonic Mach number $m_s= u_\perp/C_s$ and magnetic shear $b^2=(B_\parallel/B_\perp)^2$ are computed in the pre-shock region. Of particular interest are the relative variations of the toroidal magnetic field component $\delta B_\phi= B_{\phi 2}/B_{\phi 1} -1$ and the plasma angular velocity $\delta \Omega= \Omega_2/\Omega_1 -1$, as well as the total deflection angle of the poloidal magnetic surface $\delta i= i_2-i_1$ where $\tan i= B_t/B_\perp$, which read
 \begin{eqnarray}
\delta B_\phi &=&  (\chi -1) \frac{m^2}{m^2 - \chi}   \nonumber \\
- \delta \Omega &=& \frac{\chi-1}{m^2-\chi} \frac{m^2(r^2-r_A^2)}{m^2r_A^2-r^2} \leq  \frac{\chi-1}{m^2-\chi} \\
\tan \delta i &= & \frac{m^2(\chi-1)}{m^2-\chi} \frac{\tan i_1}{1+ \chi \tan^2 i_1 \frac{m^2-1}{m^2-\chi}}  \nonumber 
\end{eqnarray}
These quantities are plotted in Fig.~\ref{fig:ReferenceSimulationMainShockRH}. The compression rate $\chi$ is the solution of the cubic polynomial equation 
\begin{equation}
 - A \chi^3 + B \chi^2 - C\chi  + D = 0
 \label{eq:r}
\end{equation}
with
\begin{eqnarray*} 
A &=& 1 + b^2 + \frac{1+\chi_o}{\Gamma m^2_s} \\ 
B &=& \chi_o(1+b^2) + 2m^2\left ( 1 +  \frac{1+\chi_o}{\Gamma m^2_s}  - b^2 \frac{\chi_o-3}{4} \right ) \\
C &=& m^2 \left( 2\chi_o  + b^2\frac{1+\chi_o}{2} + m^2 \left ( 1 +  \frac{1+\chi_o}{\Gamma m^2_s} \right ) \right ) \\
D &=& \chi_o m^4
\end{eqnarray*}
where $\chi_o= (\Gamma+1) / (\Gamma-1)$ is the maximal compression ratio for a hydrodynamic shock. Equation~\ref{eq:r} has one positive root only for an incoming super-FM flow, namely for $n_\perp= u_\perp/V_{fm,\perp}$ larger than unity. 

We are dealing here with supersonic ($m_s>>1$) and super-A ($m>>1$) cold jets, where the dominant magnetic field is the toroidal one ($b^2\simeq (B_\phi/B_\perp)^2>>1$). The FM Mach number in the normal direction writes then $n_\perp \simeq m V_{Ap,\perp}/V_{A\phi}= m/b$,
which leads to the simplified equation for $\chi$
 \begin{equation}
 \frac{\chi_o-3}{2} \chi^2 + \left (\frac{1+\chi_o}{2} + n_\perp^2 \right ) \chi - \chi_o n_\perp^2 =0 \ .
 \end{equation}
It shows that whenever jets reach a very large FM Mach number $n_\perp$, a large compression rate $\chi \simeq \chi_o$ is possible. But this is never achieved in our case. indeed, the poloidal FM Mach number $n= u_p/V_{FM,p}$ ($> n_\perp$) writes 
 \begin{equation}
n^2 = \omega_A \frac{B_{pA} r_A^2}{B_p r^2} \frac{1- 1/m^2}{1-r_A^2/r^2} \left (\frac{u_p}{\Omega_* r_A}\right)^3 \sim  \omega_A \left (\frac{u_p}{\Omega_* r_A}\right)^3
 \end{equation}
where $\omega_A= \Omega_* r_A/V_{Ap,A}$ is the fastness parameter introduced in F97 (ratio at the Alfv\'en point of the speed of the magnetic rotator to the poloidal Alfv\'en speed). For magneto-centrifugal jets like ours, with $m^2>>1, r>>r_A$ and achieving their maximal velocity $u_p \sim \sqrt{2} \Omega_* r_A$, the FM Mach number is  $n^2 \sim \omega_A$, which is larger than but of the order of unity (see also \citealt{krasnopolsky2003}). As a consequence, we expect rather weak shocks as illustrated by the small values of $\chi$ achieved along the various shocks  (see Fig.~\ref{fig:ReferenceSimulationMainShockRH}).\\ 

\begin{figure}[H]
\centering
\includegraphics[width=.98\linewidth]{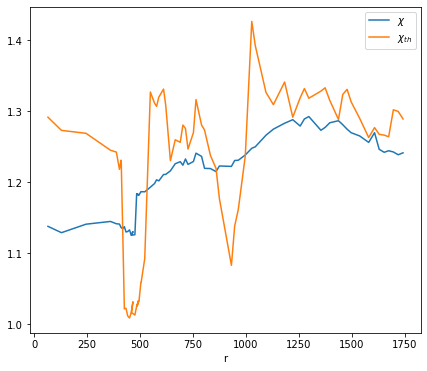}
\caption{Distribution at $t_{end}$ along the main recollimation shock of the compression ratio for simulation K2. The yellow curve is the theoretical solution $\chi_{th}$ of Eq.~\ref{eq:r}, computed using the pre-shock quantities, while the blue curve is the ratio $\chi= \rho_2/\rho_1$.} 
\label{fig:Chith}
\end{figure}

Figure~\ref{fig:Chith} plots the theoretical solution $\chi_{th}$ of Eq.~\ref{eq:r} (in yellow) computed along the main recollimation shock of our simulation K2 and compares it with the ratio $\chi= \rho_2/\rho_1$ (in blue) directly measured (see Fig.~\ref{fig:ZoomK2}). The correspondence is very good, with discrepancies remaining below a few percent. The two regions where larger differences are obtained correspond to the positions where the two smaller shocks (triggered at the spine-jet interface) merge with the main shock: the orange one near $r\sim 500$ and the cyan one near $r\sim 900$.


%
%


The shocks were detected by following all magnetic field lines anchored on the disk and looking for discontinuities. This is not obvious in a discrete grid. To do so, we computed the derivative of the toroidal magnetic field ($\delta B_\phi$) over the curvilinear abscissa along the field line, as shocks are best seen with the electric poloidal current and explored its local extrema. We checked that a different approach, based on the calculation of the refraction angle $\delta i$ of the poloidal magnetic surface, produces very similar results. This gave us the shock locations used to produce the plots in Figures \ref{fig:ReferenceSimulation} and \ref{fig:KappaShockAltitudeEvolution}. 
Since PLUTO has a shock capturing scheme, each shock is resolved and has a finite width. To determine the shock width, we checked that the density was growing within the shock as expected. Then, still following the field line, we looked for the closest local minimum and maximum in density. The positions of these extrema allowed to compute the values of the pre-shock and the post-shock quantities, respectively. Those were finally used to compute the parameters leading to the Figures \ref{fig:ReferenceSimulationMainShockRH} and \ref{fig:Chith}.

\end{appendix}
\end{document}